\documentclass[preprint,12pt]{elsarticle}

\usepackage{amssymb}
\usepackage{amsmath}

\usepackage{natbib}
\usepackage{newtxtext}
\usepackage{newtxmath}
\usepackage{natbib}
\usepackage{hyperref}
\usepackage{comment}
\usepackage{algorithm}
\usepackage{algorithmic}

\usepackage{float}

\usepackage{tabularx} 
\usepackage{booktabs}       
\usepackage{multirow}
\usepackage{diagbox}

\usepackage{subcaption}
\usepackage{wrapfig} 

\usepackage{amssymb}
\usepackage{fancyhdr}
\usepackage[version=4]{mhchem}
\usepackage{titlesec}
\hypersetup{
    colorlinks=true,
    linkcolor=blue,  
}

\usepackage{graphicx}
\usepackage{adjustbox}

\renewcommand{\ref}[1]{\textcolor{blue}{\hyperref[#1]{\nameref*{#1}}}}

\journal{Computer Physics Communications}

\begin{document}

\begin{frontmatter}



\title{An efficient, adaptive solver for accurate simulation of multicomponent shock-interface problems for thermally perfect species}
 
\author[inst1]{Yuqi Wang}  
\author[inst2]{Ralf Deiterding}
\author[inst1]{Jianhan Liang\corref{cor1}}
\cortext[cor1]{Correponding author: jhleon@vip.sina.com}

\affiliation[inst1]{organization={Hypersonic Technology Laboratory, National University of Defense Technology},
                city={Changsha},
                postcode={410073}, 
                country={P.R.China}}
                
\affiliation[inst2]{organization={AMROC CFD}, 
                addressline={Brookweg 167}, 
                city={Oldenburg}, 
                postcode={26127}, 
                country={Germany}}

\begin{abstract}
A second-order-accurate  finite volume method, hybridized by blending an extended double-flux algorithm and a traditionally conservative scheme, is developed. In this scheme, hybrid convective fluxes as well as hybrid interpolation techniques are designed to ensure stability and accuracy in the presence of both material interfaces and shocks.
Two computationally efficient approaches, extended from the original double-flux model, are presented to eliminate the well-known "pressure oscillation" phenomenon at material interfaces observed with the traditional conservative scheme. Numerous verfication simulations confirm that the method is capable of handling multi-dimensional shock-interface problems reliably and efficiently, even in the presence of viscous and reactive terms.
\end{abstract}


\begin{highlights}
\item Extension of the the original double-flux model to multicomponent mixtures with thermally perfect species to only one auxillary variable required and with better physical interpretability.
\item Development of a new hybrid method by blending an extended double-flux algorithm and a standard conservative scheme by utilizing hybrid convective fluxes and interpolation techniques.
\item Verification and validation of the presented new hybrid solver by extensive one- and multi-dimensional benchmark tests and also some very complex chemically reacting problems
\end{highlights}


\end{frontmatter}



\section{Introduction}
\label{sec1}
It is nowadays well established that a fully conservative scheme will usually lead to spurious pressure and velocity oscillations at material interfaces with variable specific heat ratio $\gamma$ when solving the Euler or Navier-Stokes equations in the hyperbolic regime~\cite{larrouturou1991preserve,karni1994multicomponent,abgrall1996prevent,ton1996improved}.
Earlier works by Karni \cite{karni1994multicomponent} and Jenny et al. \cite{jenny1997correction} recognized that this pressure oscillation problem is rooted in the numerical averaging or smearing of $\gamma$ across cells.
To eliminate this issue, one way is to develop "sharp-interface" methods, which essentially avoid smearing out $\gamma$ by tracking the propagating material front~\cite{ton1996improved,shyue1998efficient,johnson2020conservative}.
One classic method among them was presented by Karni \cite{karni1994multicomponent} who evolved the original governing equations into a non-conservative formulation by solving additionally the advection equation of one primitive variable.
Using this approach, artificial pressure oscillations were perfectly cured and Abgrall \cite{abgrall1996prevent} later extended this approach to handle strong shocks as well.
Besides, there is also a family of non-oscillatory approaches named as the "Ghost Fluid Method" (GFM) \cite{fedkiw1997high,fedkiw1998efficient,fedkiw1999non,fedkiw2002coupling}, which preserve the interface sharpness by defining a level-set function and using ghost fluid cells to swap information across the interface.
The GFM methods combine the advantages of both Eulerian and Langrangian approaches and have been widely used to handle multi-phase fluid problems~\cite{chang2013direct,luo2015conservative,xu2022interface}.

Another approach is to control the pressure fluctuation by artificially remedying the solution of the conservative total energy. One of the typical strategies was put forward by Jenny et al. \cite{jenny1997correction} who implemented a simple correction onto the total energy per volume, which preserved the interface free of oscillations. Although during this process the scheme conservation is inevitably lost, numerical stability and convergence issues have not been reported~\cite{jenny1997correction,billet2003adaptive,houim2011low,lv2014discontinuous}.

However, all the methods mentioned above were tailored for calorically perfect gases whose thermodynamic properties, like $C_{p}, C_{v}$ and $\gamma$, do not change intricately with the local temperature, and thus formerly mentioned strategies, like discretizing an additional primitive equation or simple energetic correction, could work. Note that for calorically perfect gases, it is generally not necessary to consider the possible issue of the schemes in the presence of shock waves since the specific heat ratio $\gamma$ in this model remains constant across shocks .

In order to consider species with temperature-dependent  thermodynamic properties, Abgrall proposed the original "double-flux" model \cite{abgrall2001computations}, which was subsequently improved by Billet and Abgrall \cite{billet2003adaptive}. This approach basically followed the strategy of a non-conservative energetic correction procedure but was explicitly tailored for thermally perfect gases. The concept was to define two sets of numerical fluxes at every edge, for respectively its left- and right-adjacent cell such that a cell could evolve purely based upon its own thermodynamic properties, i.e. the thermodynamic properties got effectively frozen during the cell-based time evolution.
Apparently, the original double-flux model was non-conservative and could not handle strong shocks directly because inconsistent fluxes do not guarantee convergence to the correct weak solution. The methodology of the original double-flux model as well as its deficiency will be further discussed later in Section \ref{sec3.1}.

The double-flux approach by Billet and Agbrall \cite{billet2003adaptive} has been used subsequently to simulate complex, real-world mixing and combustion problems, as done in \cite{houim2011low,houim2016role,goodwin2016effect}.
In recent years, there have been also some efforts to further develop the original method.
For example, Houim and Kuo \cite{houim2011low} extended the original method of second order accuracy to the high-order WENO scheme. Lv and Ihme \cite{lv2014discontinuous} as well as Johnson and Kercher \cite{johnson2020conservative} extended it separately to the Discontinuous Galerkin (DG) framework for solving chemically reacting Navier-Stokes equations. The compatibility of the double-flux model with the DG method in the presence of viscous terms has recently been analyzed \cite{bando2020comparison}. It has also been reported that the oscillations in addition create spurious temperature spikes and species errors \cite{johnsen2012preventing,lv2013development}, which may lead to significant numerical artifacts. 

In this paper, based on the double-flux model, we design a hybrid solver containing both hybrid fluxes and hybrid interpolations, and thus ensure convergence towards the correct weak solution.
Specifically, an extended double-flux scheme is developed and used in regions where the pressure is smooth, including the material interfaces, while a traditionally conservative scheme is implemented especially for shock capturing.
For hybridization switching, many works have contributed to the accurate and efficient recognition of discontinuities within smooth solutions by using an explicit shock sensor \cite{qiu2005comparison}-\cite{krivodonova2004shock}. A comprehensive comparison and summary can be found in \cite{li2010hybrid}.
For example, Hill and Pullin \cite{hill2004hybrid} developed and combined the tuned central difference (TVD) scheme with the WENO method for compressible large-eddy simulation.
In their hybridization methdology, to limit the usage of WENO scheme to shocks, a  scaled criterion of the smoothness indicator over the WENO candidate stencils was employed for determination of the nodal flux $\hat{f}_{j+1/2}$.
Ziegler et al. \cite{ziegler2011adaptive} then improved the shock-detection technique by applying the Liu entropy condition calculated by a Roe-averaged approximate Riemann solver.  
Besides that, instead of explicitly switching between the central-difference and upwind schemes, Jiang et al \cite{jiang2016hybrid} used a multi-dimensional optimal order detection (MOOD) approach \cite{clain2011high,loubere2014new,dumbser2014posteriori} by distinguishing the "problematic" cells with the usage of physical and numerical admissibility conditions in order to implement in these cells an upwind discretization.
In the present study, since both of the utilized fluxes  are calculated based on a monotonicity preserving upwind scheme, the choice of shock sensor is less crucial for stable, accurate shock-interface simulation.
Therefore, the normalised second derivative of pressure, that has been used in the classical artificial viscosity method \cite{davis1987simplified}, is employed.

Our codes are verified to be capable of handling complex scenarios with the presence of both material interfaces and shock discontinuities, as well as their complicated interaction.
Generally, our improved strategy is more efficient and straightforward to implement than enlarging the whole equation system and requires just a minor modification on the total energy and minimal additional coding efforts.

\section{Governing equations and numerical methodology}
\label{sec2}
\subsection{Chemically reacting compressible Navier-Stokes equations}
\label{sec2.1}
In this paper, we consider finite volume discretizations of the compressible multicomponent Navier-Stokes system to describe the multi-dimensional reacting flows with detailed chemical kinetics. 
Soret and Dufour effects, external body forces (e.g. the gravity) and radiant heat transfer are all neglected. 
The basic equation reads
\begin{gather}
\label{eq1}
    \frac{\partial \boldsymbol{U}}{\partial t} + \frac{\partial \boldsymbol{F}}{\partial x} + \frac{\partial \boldsymbol{G}}{\partial y} = \frac{\partial \boldsymbol{F^{v}}}{\partial x} + \frac{\partial \boldsymbol{G^{v}}}{\partial y} + \boldsymbol{S_{\text{chem}}} \;,
\end{gather}
where the state vector $\boldsymbol{U}$ of the conservative variables and the convective flux vectors $\boldsymbol{F}$, $\boldsymbol{G}$ read
\begin{gather}
    \boldsymbol{U} = \begin{pmatrix}
        \rho_1 \\
        \vdots \\
        \rho_{N_{sp}} \\
        \rho u \\
        \rho v \\
        \rho E
    \end{pmatrix}, \quad
    \boldsymbol{F} = \begin{pmatrix}
        \rho_1 u \\
        \vdots \\
        \rho_{N_{sp}} u \\
        \rho u^2 + p \\
        \rho u v \\
        (\rho E + p)u
    \end{pmatrix}, \quad
    \boldsymbol{G} = \begin{pmatrix}
        \rho_1 v \\
        \vdots \\
        \rho_{N_{sp}} v \\
        \rho u v \\
        \rho v^2 + p \\
        (\rho E + p)v
    \end{pmatrix}
\end{gather}
and the viscous fluxes $\boldsymbol{F^{v}}$, $\boldsymbol{G^{v}}$ read
\begin{align}
    \boldsymbol{F^{v}}=
        \begin{pmatrix}
        -J_{x,1} \\
        ... \\
        -J_{x,N_{sp}} \\
        \tau_{xx} \\
        \tau_{xy} \\
        u\tau_{xx} + v\tau_{xy} - q_x\\
        \end{pmatrix}, \quad &
    \boldsymbol{G^{v}}=
        \begin{pmatrix}
        -J_{y,1} \\
        ... \\
        -J_{y,N_{sp}} \\
        \tau_{yx} \\
        \tau_{yy} \\
        u\tau_{yx} + v\tau_{yy} - q_y\\
        \end{pmatrix}.
\end{align}
Here, $\boldsymbol{J_{k}}$ denotes the molecular diffusion flux vectors of species $k$ with components
\begin{equation}
\begin{aligned}
   J_{x,k} &= \rho Y_{k}\left[\left(-\frac{1}{X_{k}}\right) D_{k}\left(\frac{\partial X_i}{\partial x}+ \frac{X_{k}-Y_{k}}{p}\frac{\partial p}{\partial x}\right)\right], \\
   J_{y,k} &= \rho Y_{k}\left[\left(-\frac{1}{X_{k}}\right) D_{k}\left(\frac{\partial X_i}{\partial y}+ \frac{X_{k}-Y_{k}}{p}\frac{\partial p}{\partial y}\right)\right].
\end{aligned}
\end{equation}
The viscous stress tensor $\boldsymbol{\tau}$ is given by 
\begin{equation}
\begin{aligned}
   \tau_{xx} &= \mu \left(2\frac{\partial u}{\partial x}-\frac{2}{3} \left(\frac{\partial u}{\partial x}+\frac{\partial v}{\partial y}\right)\right), \\
   \tau_{xy} &= \mu \left(\frac{\partial u}{\partial y}+\frac{\partial v}{\partial x}\right), \\
   \tau_{yy} &= \mu \left(2\frac{\partial v}{\partial y}-\frac{2}{3} \left(\frac{\partial u}{\partial x}+\frac{\partial v}{\partial y}\right)\right)
\end{aligned}
\end{equation}
and the heat flux vector $\boldsymbol{q}$ is determined by 
\begin{align}
   q_x = \sum_{k=1}^{N_{sp}} J_{x,k}h_{k} - \lambda\frac{\partial T}{\partial x},  \quad & 
   q_y = \sum_{k=1}^{N_{sp}} J_{y,k}h_{k} - \lambda\frac{\partial T}{\partial y},
\end{align}
where $h_{k}$ is the specific enthalpy of the $k$-th species and its contribution to the total enthalpy $h$ is expressed as
\begin{equation}
    h = \sum_{k=1}^{N_{sp}} Y_{k}h_{k}.
\end{equation}

Mixture properties are determined from the pure species properties and are re-evaluated every numerical time step.
The molecular diffusion coefficients $D_k$ are obtained with the multi-species diffusion model considering the pressure effect
\begin{equation}
    D_k = \frac{1-X_k}{\sum_{j\neq k}^{N_{sp}}(X_k/D_{kj})}\frac{p_0}{p},
\end{equation}
where $p_0$ is the ambient pressure.
The mixture-averaged viscosity $\mu$ is given by the semi-empirical formula due to Wilke \cite{wilke1950viscosity} and modified by Bird et al \cite{bird1961transport}.
For the mixture-averaged thermal conductivity $\lambda$, a combination averaging formula \cite{mathur1967thermal} was used 
\begin{equation}
    \lambda = \frac{1}{2}\left(\sum_{k=1}^{N_{sp}} X_k\lambda_k + \frac{1}{\sum_{k=1}^{N_{sp}} X_k/\lambda_k}\right).
\end{equation}
The reaction source term $\boldsymbol{S_{\text{chem}}}$ in Eq. \ref{eq1} reads 
\begin{equation}
\boldsymbol{S_{\text{chem}}}=\left(\dot{\omega}_{1}, \ldots, \dot{\omega}_{N_{sp}},0,0,0\right)^{T}.
\end{equation}
The chemical production rates $\dot{\omega}_{k}\left(\rho_{1}, \ldots, \rho_{N_{sp}}, T\right)$ are derived from a reaction mechanism that consists of $J$ chemical reactions
\begin{equation}
\sum_{k = 1}^{N_{sp}} \nu_{j k}^{f} S_{k} \rightleftharpoons \sum_{k = 1}^{N_{sp}} \nu_{j k}^{r} S_{k}, \quad j = 1, \ldots, J,
\end{equation}
where $\nu_{j i}^{f}$ and $\nu_{j i}^{r}$ are the stoichiometric coefficients of species $S_k$ appearing as a reactant and as a product. The entire molar production rate of species $S_k$ per unit volume is then given by
\begin{equation}
\dot{\omega}_{k} = \sum_{j = 1}^{J}\left(\nu_{j k}^{r}-\nu_{j k}^{f}\right)\left[k_{j}^{f} \prod_{k = 1}^{N_{sp}} \left(\frac{\rho_k}{W_k}\right)^{\nu_{j k}^{f}}-k_{j}^{r} \prod_{k = 1}^{N_{sp}}\left(\frac{\rho_k}{W_k}\right)^{\nu_{j k}^{r}}\right], \quad k = 1, \ldots, N_{sp},
\end{equation}
with $k_{j}^{f}(T)$ and $k_{j}^{r}(T)$ denoting the forward and backward reaction rate of each chemical reaction, respectively.
Each forward reaction rate is given by an Arrhenius formula and the backward reaction rates is calculated from the respective chemical equilibrium constant. In the present study, a detailed chemical model with 13 species and 27 reversible reaction steps developed by Burke et al. \cite{burke2012comprehensive} was employed.

For closure of the above system, we have the ideal gas law which reads
\begin{equation}
    p = \rho R T = \sum_{k=1}^{N_{sp}} \rho Y_k \frac{RU}{W_k} T,
\end{equation}
where $R$ is the gas constant and $W_k$ is the relative molecular mass of species $k$. 

A thermally perfect equation of state is used to determine the total energy per volume $E$ of gas with variable specific heat ratio $\gamma$ dependent on the temperature as
\begin{equation}
\begin{aligned}
\label{eq12}
    E &= h - \frac{p}{\rho} + \frac{1}{2} \left({u}^2+{v}^2\right) \\
    &= \sum_{k=1}^{N_{sp}} \left( Y_k h^f_{k0} + Y_k \int_{T_0}^T C_{pk}(s)ds \right) - \frac{p}{\rho} + \frac{1}{2} \left({u}^2+{v}^2\right).
\end{aligned}
\end{equation}
This property is of vital importance for solving chemical non-equilibrium flows, which are typically exposed to large temperature gradients. 
The specific heat capacity at constant pressure, $C_{pk}$, as well as the partial enthalpy $h_k$ of the $k$-th species are determined by the CHEMKIN library.

In contrast, we also write down the equation of state for a calorically perfect gas
\begin{equation}
\label{eq13}
    E_{\text{cal}} = \frac{p}{\gamma-1} + \frac{1}{2} \left({u}^2+{v}^2\right).
\end{equation}

\subsection{Numerical schemes}
\label{sec2.2}
In this section, we present the numerical methodology for solving the multi-dimensional chemically reacting Navier-Stokes equations.
An overall second-order-accurate Strang splitting method is employed for decoupling of the transport and reactive parts and thereby ensure computational efficiency and robustness, given by 
\begin{align}
\boldsymbol{U}^{n+1}=\mathcal{S}^{\left(\frac{1}{2} \Delta t\right)} \mathcal{H}^{(\Delta t)} \mathcal{S}^{\left(\frac{1}{2} \Delta t\right)}\left(\boldsymbol{U}^{n}\right),
\end{align}
where $\mathcal{S}$ and $\mathcal{H}$ stand for the time operator of the inhomogeneous ordinary differential part and the homogeneous partial differential part, respectively.

For the homogeneous part, an unsplit finite-volume approach is employed with each volume advanced by accounting for all flux contributions in a single step.
In each dimension, a second-order-accurate MUSCL (Monotone Upstream-centered Schemes for Conservation Laws) scheme is employed for discretization of the convective terms and a second-order-accurate central difference scheme for the viscous terms.
In the MUSCL reconstruction, determination of the edge value $\boldsymbol{U}_{j\pm1/2}$ is first achieved by a linear interpolation between the three cells $j-1$, $j$ and $j+1$.  
Special factors $\Phi_{j\pm1/2}^{\pm}$, called the slope limiters, are utilized to reconstruct the gradients used for this interpolation to ensure the TVD property and positivity of the species mass fractions:
\begin{subequations}
\begin{align}
\tilde{U}_{j+\frac{1}{2}}^{L}=U_{j}^{l}+\frac{1}{4}\left[(1-\omega) \Phi_{j-\frac{1}{2}}^{+} \Delta_{j-\frac{1}{2}}+(1+\omega) \Phi_{j+\frac{1}{2}}^{-} \Delta_{j+\frac{1}{2}}\right], \\
\tilde{U}_{j-\frac{1}{2}}^{R}=U_{j}^{l}-\frac{1}{4}\left[(1-\omega) \Phi_{j+\frac{1}{2}}^{-} \Delta_{j+\frac{1}{2}}+(1+\omega) \Phi_{j-\frac{1}{2}}^{+} \Delta_{j-\frac{1}{2}}\right].
\end{align}    
\end{subequations}

In this present paper, the minmod limiter is applied to ensure calculation stability.
Note, that only for $\omega= 0$ the following conservation property is satisfied:
\begin{eqnarray}
U_{j}^{l} & = & \frac{1}{\Delta x} \int_{x_{j-1 / 2}}^{x_{j+1 / 2}} \tilde{U}(\xi) d \xi
\end{eqnarray}

High-order spatial methods are unstable under the forward-Euler time integration method.
The method of lines is therefore a frequently selected approach for the temporal discretization of hyperbolic problems.
In this paper, an explicit two-stage strong stability preserving Runge-Kutta method (SSPRK2) of second order time accuracy is employed. Its steps are
\begin{subequations}
\begin{align}
    K_1 &= \Delta tS(t^n, \boldsymbol{U}^n) ,  \\
    K_2 &= \Delta tS(t^n+\Delta t, \boldsymbol{U}^n+K_1), \\
    \boldsymbol{U}^{n+1} &= \boldsymbol{U}^n + \frac{1}{2}[K_1+K_2].
\end{align}    
\end{subequations}

The Riemann solver used in this paper is the HLLC solver \cite{toro2013riemann}, which inherently preserves the velocity equilibrium across the material interface after one time step $dt$. The latter is crucially important for developing a new method based on the double-flux model, which will be seen in later sections.
Other flux solvers with this property, such as the AUSM$^+$ scheme \cite{liou1996sequel}, could also be applied.
The viscous diffusion terms are evaluated simply by standard second-order-accurate central differences.
Each integration of the source term is done with the fourth-order semi-implicit Runge-Kutta method (GRK4A) \cite{kaps1979generalized}, which is capable of handling the stiffness of the ODEs due to the detailed chemical kinetics.

\section{Extended double-flux method}
\label{sec3.1}
Spurious oscillations will be generated at a material interface when solving the multicomponent Euler or Naiver-Stokes equations with variable specific heat ratio $\gamma$. The double-flux algorithm is an efficient method that is able to preserve the pressure and velocity across a material interface. The approach has been proposed by Abgrall et al. \cite{abgrall1996prevent,billet2003adaptive} and was later extended to a high-order WENO scheme \cite{houim2011low} in the finite difference framework and a discontinuous Galerkin (DG) method \cite{lv2014discontinuous}, respectively. According to the authors' knowledge, it is the only method that can handle multicomponent flows with arbitrary number of species whose $C_{pi}$ depend on the temperature. 

\subsection{Methodology}
The original double-flux model required the two auxillary variables, $\gamma$ and $\rho h_{0}^{m}$ and was tailored for the JANAF thermodynamic tables.
In this section, we improve the algorithm by requiring only one auxillary variable and nevertheless obtain perfect pressure and velocity equilibrium at a material front. Furthermore, we modified the new algorithm to be applicable to the NASA library.
Two approaches of the extended double-flux method are presented and compared in Sections \ref{sec3.1.1} and \ref{sec3.1.2}, respectively.

To begin with, a material front can be analyzed by using the one-dimensional inviscid Euler equations
\begin{subequations}
\label{eq21}
\begin{align}
\frac{\partial (\rho Y_{i})}{\partial t} & = - \frac{\partial (\rho Y_{i} u)}{\partial x} \label{eq21a}, \\
\frac{\partial (\rho u)}{\partial t} & = - \frac{\partial (\rho  u^2 + p)}{\partial x} \label{eq21b}.
\end{align}
\end{subequations}
Since $u, p$ are constant across a material front, Eq. \ref{eq21} can be reduced to 
\begin{subequations}
\label{eq22}
\begin{align}
\label{eq22a}
\delta\left(\rho Y_{i}\right) & = -\sigma u \Delta\left(\rho Y_{i}\right), \\
\label{eq22b}
\delta(\rho u) & = -\sigma u^{2} \Delta \rho,
\end{align}
\end{subequations}
where $\delta$ and $\Delta$ denote the temporal and spatial difference, respectively, and $\sigma = {dt}/{dx}$.
Summing Eq. \ref{eq22a} from $i$ = 1 to $N$ together and substituting the result into Eq. \ref{eq22b} will yield the relation
\begin{align}
u^{n+1} = u^{n}.
\end{align}
The latter means that the normal velocity $u$ is inherently preserved after one time step before it is contaminated by the pressure fluctuation. 

Substituting Eq. \ref{eq12} into the energy equation (where the pressure oscillation originates), we have
\begin{equation}
\begin{aligned}
\label{eq24}
\delta\left(\sum \rho Y_{i} h_{i 0}^{f}\right) + \delta\left(\sum \rho Y_{i} \int_{T_{0}}^{T} C_{p i}(s) ds\right) - \delta\left(p\right) + \delta\left(\frac{1}{2} \rho u^{2}\right) = \\
-\sigma u \left(\Delta\left(\sum \rho Y_{i} h_{i 0}^{f}\right) + \Delta\left(\sum \rho Y_{i} \int_{T_{0}}^{T} C_{p i}(s) ds\right) - \Delta\left(p\right) + \Delta\left(\frac{1}{2} \rho u^{2}\right)\right).
\end{aligned}
\end{equation}

\begin{figure}[H]
    \centering
    \includegraphics[width=0.4\textwidth]{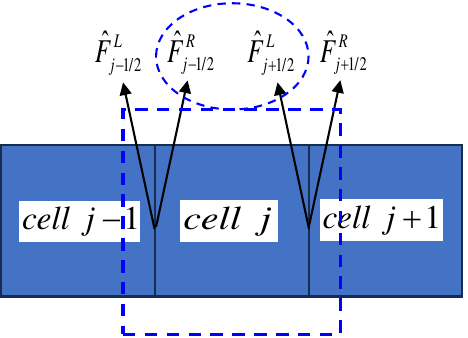}
    \caption{Schematic diagram of cell-based implementation of the double-flux type method.}
    \label{fig3-1-1}
\end{figure}

Due to the presence of the integral term of $C_{pi}(s)$, directly analyzing Eq.~\ref{eq24} is challenging.
A better way is to first reformulate Eq.~\ref{eq24} by transforming the equation of state of Eq. \ref{eq12} into a familiar expression like Eq. \ref{eq13}.
In the original work by Billet and Abgrall \cite{billet2003adaptive}, they decomposed the integral $\int_{T_0}^T C_{pi}(s)ds$, i.e., the sensible enthalpy. Each $C_{pi}$ is dependent on the temperature and obtained from the JANAF tables by piecewise linear interpolation. 
The integral of $C_{pi}$ except from the last temperature interval can be merged into the enthalpy of formation $h_{i0}^f$.
Then the total enthalpy per volume can be written as
\begin{equation}
\begin{aligned}
h_{i} & = h_{i 0}^{f}+\sum_{k = 1}^{p} \int_{T_{k-1}}^{T_{k}}\left(\alpha_{i}^{k} s+\beta_{i}^{k}\right) \mathrm{d} s+\int_{T_{p}}^{T}\left(\alpha_{i}^{p} s+\beta_{i}^{p}\right) \mathrm{d} s = h_{i 0}^{m}+\int_{T_{p}}^{T}\left(\alpha_{i}^{p} s+\beta_{i}^{p}\right) \mathrm{d} s,
\end{aligned}
\end{equation}
where $\alpha_{i}^{k}$, $\beta_{i}^{k}$ are the constants of piecewise linear interpolation.  Introducing  
\begin{align}
\label{eq201}
\overline{C_{pi}} & = \frac{a_{i}^{p}}{2}\left(T+T_{p}\right)+b_{i}^{p}
\end{align}
as the equivalent heat capacity at constant pressure of species $i$ at the last temperature interval $\left[T_{p}, T\right]$ and using
\begin{equation}
\begin{aligned}
\label{eq25}
    \hat{h}_{i 0}^{m} & = h_{i 0}^{m}-\overline{C_{pi}} T_{p}, \quad     \overline{\gamma} &= \frac{\overline{C_{p}}}{\overline{C_{p}}-R}
\end{aligned}
\end{equation}
the total energy per volume $E$ can be reformulated into
\begin{equation}
\begin{aligned}
\label{eq199}
E & = h_{0}^{m}+\sum_{i = 1}^{N} \overline{C_{pi}} Y_{i} (T-T_{p})-\frac{p}{\rho}+\frac{u^{2}}{2} = \hat{h}_{0}^{m}+\sum_{i = 1}^{N} \overline{C_{pi}} Y_{i} T-\frac{p}{\rho}+\frac{u^{2}}{2} \\
& = \hat{h}_{0}^{m}+\overline{C_{p}} T-\frac{p}{\rho}+\frac{u^{2}}{2} = \hat{h}_{0}^{m}+\frac{p}{\rho(\overline{\gamma}-1)}+\frac{u^{2}}{2}.
\end{aligned}
\end{equation}
Substituting Eq. \ref{eq199} into the energy equation of the hyperbolic system, we then have
\begin{equation}
\begin{aligned}
&\delta\left(\rho \hat{h}_{0}^{m}\right) + p \delta\left(\frac{1}{\overline{\gamma}-1}\right) + \frac{1}{\overline{\gamma}-1} \delta p + \delta\left(\frac{1}{2} \rho u^{2}\right) \\
=- &\sigma u \biggl[
\Delta\left(\rho \hat{h}_{0}^{m}\right) + p \Delta\left(\frac{1}{\overline{\gamma}-1}\right) +  \left(\frac{1}{\overline{\gamma}-1}\right)\Delta p + \Delta\left(\frac{1}{2} \rho u^{2}\right)
\biggr].
\end{aligned}
\label{eq200}
\end{equation}
In Eq. \ref{eq200}, we can analyze that at a material front to avoid pressure fluctuations, one would have to fix the auxillary variables $\rho \hat{h}_{0}^{m}$ and $\overline{\gamma}$ together during the discrete cell marching.
In the specific implementation, the gradients of the hyperbolic numerical fluxes at the iterative computation of cell $j$ in each dimension $i$ are thus calculated by
\begin{equation}
    \frac{\partial F}{\partial x_i} = \frac{\hat{F}\left(\boldsymbol{U}_{j+1/2}^L,\boldsymbol{U}_{j+1/2}^R,\hat{\gamma}_j,(\rho \hat{h}_{0}^{m})_j\right) - \hat{F}\left(\boldsymbol{U}_{j-1/2}^L,\boldsymbol{U}_{j-1/2}^R,\hat{\gamma}_j,(\rho \hat{h}_{0}^{m})_j\right)}{\Delta x_i}.
\end{equation}
This implementation is illustrated in Fig. \ref{fig3-1-1}. Due to the potential inconsistency between the numerical fluxes $\hat{F}_{j-1/2}^L$ and $\hat{F}_{j-1/2}^R$, non-conservation is inevitably found for the energy term.

However, it can be seen that the derivation of the original double-flux model is tedious, mostly attributed to its complex modelling of the specific energy $E$. 
Although the pressure equilibrium is preserved, one obvious flaw of this method is that the reconstruction of "mixed enthalpy of formation" $\hat{h}_{0}^{m}$ is void of any physical basis.
In the implementation, this method also requires two thermodynamic parameters $\overline{\gamma}$ and $\rho \hat{h}_{0}^{m}$ to be frozen, which complicates the programming and might bring more conservation loss to the energy term.

Having realized the deficiencies of original double-flux model, we found that we could make some improvement to it.
In this paper, we modified the algorithm to require only one auxillary variable $\hat{\gamma}$ to eliminate pressure oscillation, by altering the reformulation model of the energy term and greatly simplifying the derivation process. 
Furtheron, we extended the algorithm to mixtures for which the specific heats $C_{pi}$ are obtained from the NASA database by polynomial interpolation, making it more trivial and straightforward to be used with the CHEMKIN library which has been adopted in most large-scale computational fluid dynamic codes.
The concept we employed here is to directly average the enthalpy-based part in the total energy per volume $E$.
By utilizing this concept, only one fixed auxillary variable instead of two will be required.
We also succeed in preserving the physical fundamentals better since $\hat{\gamma}$ is nothing but the averaged specific heat ratio of the enthalpy-based part.

\subsubsection{Approach A}
\label{sec3.1.1}

One approach is to directly reformulate the sensible enthalpy contribution in the specific enthalpy, independent of the enthalpy of formation, by introducing $\hat{C}_{pi}$ as an averaged specific heat capacity at constant pressure of species $i$  as
\begin{equation}
\begin{aligned}
\label{eq25}
    \hat{C}_{pi} &= \frac{\int_{T_{\text{ref}}}^{T}C_{pi}(s)ds}{T}, \quad 
    \hat{C}_{p} &= \sum_{i=1}^{N} \hat{C}_{pi}Y_{i}, \quad     \hat{\gamma} &= \frac{\hat{C_{p}}}{\hat{C_{p}}-R},
\end{aligned}
\end{equation}
where $T_{\text{ref}}$ is an arbitrary reference temperature except for 0 K, which is sensible as the polynomial representations are usually not valid down to absolute zero. $\hat{C}_{p}$ is the averaged specific heat capacity at constant pressure and $\hat{\gamma}$ is the averaged specific heat ratio.
Note that $\hat{C}_{pi}$ is different from the equivalent heat capacity $\overline{C}_{pi}$ defined in Eq. \ref{eq201} as the former one has specific physical meaning.
The enthalpy of formation $h_{i0}^f$ defined at $T_{\text{ref}}$, on the other hand, is left unchanged.

By doing such, rigorous splitting between the temperature-independent enthalpy of formation and temperature-dependent sensible energy is ensured, and  Eq. \ref{eq12} can then be reformulated into a new form resembling the calorically perfect gases, i.e., the ideal gas equation with constant $C_{p}$ and $\gamma$:
\begin{equation}
\label{eq27}
    E = h_{0}^{f} - \frac{1}{\hat{\gamma}-1}\frac{p}{\rho} + \frac{1}{2}({u}^2+{v}^2)
\end{equation}
Reformulating and extending Eq. \ref{eq24} using the last equation, we obtain 
\begin{equation}
\begin{aligned}
&\delta\left(\rho h_{0}^{f}\right) + p \delta\left(\frac{1}{\hat{\gamma}-1}\right) + \frac{1}{\hat{\gamma}-1} \delta p + \delta\left(\frac{1}{2} \rho u^{2}\right) \\
= -&\sigma u \biggl[
\Delta\left(\rho h_{0}^{f}\right) + p \Delta\left(\frac{1}{\hat{\gamma}-1}\right) + \left(\frac{1}{\hat{\gamma}-1}\right) 
\Delta p  + \Delta\left(\frac{1}{2} \rho u^{2}\right)
\biggr].
\end{aligned}
\label{eq28}
\end{equation}

The conservative enthalpy of formation $\rho h_{0}^{f}$ is mathematically a convex combination of the partial density $\rho Y_i$ given by 
\begin{equation}
\begin{aligned}
\label{eq29}
    \left(\rho h_{0}^{f}\right)_t = -\sum_{i=1}^K \left( h_{i0}^{f} \sigma u (\rho Y_i)_x \right) = -\sigma u\sum_{i=1}^K \left( \rho Y_ih_{i0}^{f} \right)_x = -\sigma u\left(\rho h_{0}^{f}\right)_x.
\end{aligned}
\end{equation}

Combining Eqs. \ref{eq22} and \ref{eq29}, the terms regarding the conservative formation of enthalpy and kinetic energy on both sides of Eq. \ref{eq28} inherently cancel each other out. 
Now we have just to freeze $\hat{\gamma}$  during the cell-based evolution $\left[t_n,t_{n+1}\right]$ to obtain $\delta p=0$. Since we already have $\Delta p=0$, and thus the equilibrium across a material interface will be preserved. 
Compared to the original double-flux method, the advantage of this new approach is evident: it requires only one auxiliary variable $\hat{\gamma}$, instead of two auxillary variables, $\rho h_{0}^{m}$ and $\gamma$, when modifying the cell-based energy fluxes.
This allows for a more concise mathematical derivation and programming procedure.

\subsubsection{Approach B}
\label{sec3.1.2}
Besides the $\textit{Approach A}$, we also present in this paper another more attractive approach by directly modelling the specific enthalpy $h$, so that Eq. \ref{eq12} can be reformulated into a form similar to Eq. \ref{eq13} by defining
\begin{equation}
\label{eq31}
    \hat{C}_{p} = \frac{h}{T}, \quad 
    \hat{C}_{p} = \sum_{i=1}^{N} \hat{C}_{pi}Y_{i}, \quad     \hat{\gamma} = \frac{\hat{C_{p}}}{\hat{C_{p}}-R}.
\end{equation}
Eq. \ref{eq12} can then be reformulated into 
\begin{equation}
\label{eq32}
    E = \hat{E}_{\text{cal}}  \equiv  \frac{1}{\hat{\gamma}-1}\frac{p}{\rho} + \frac{1}{2}({u}^2+{v}^2).
\end{equation}
In this approach, we no longer need to manually specify an auxillary constant temperature $T_{\text{ref}}$, which significantly optimized the programming procedure and the issue of non-physical calculation of $\hat{C}_{p}$ got avoided from the root. 
By conducting a derivation analogous to the one of Eq. \ref{eq28}, one can show that the equilibrium of velocity and pressure is strictly guaranteed if $\hat{\gamma}$ is frozen during the cell update.

\subsection{Verification and validation}
\label{sec3.2m}
\subsubsection{Correctness and efficiency}

To verify our presented method and also the correctness of the implementation, we first test the convergence of the extended double-flux method by solving the inviscid Euler equations. However, for thermally perfect gases, the choice of a suitable exact solution is a non-trivial task since the variation of an arbitrary quantity would generally result in variations in $\Delta \gamma$ due to the density, pressure, temperature as well as the species mass fractions coupled by the ideal gas equation.
Thus, we turn to an exact solution derived from a problem with trivial variation of $\gamma$ 
\begin{equation}
\begin{array}{c}
\rho = h(x-u t)+\rho_{\infty}, \\
u = u_{\infty}, \\
p = p_{\infty},
\end{array}
\end{equation}
with $h\left(\psi\right)$ being an arbitrary smooth function. Here, we choose a cosine function.

The entire domain ranges from 0 to 1 and the periodic boundary condition (BC) is employed.
In this one-dimensional solution, the wave does not deform because the speed $u$ is constant. This is also known as an entropy wave, which is a linear degenerated wave. In reality, it relates to the contact surfaces or shear layers with only variations in density and constant velocity and pressure.

\begin{figure}[H]
    \centering
    \includegraphics[width=0.65\textwidth]{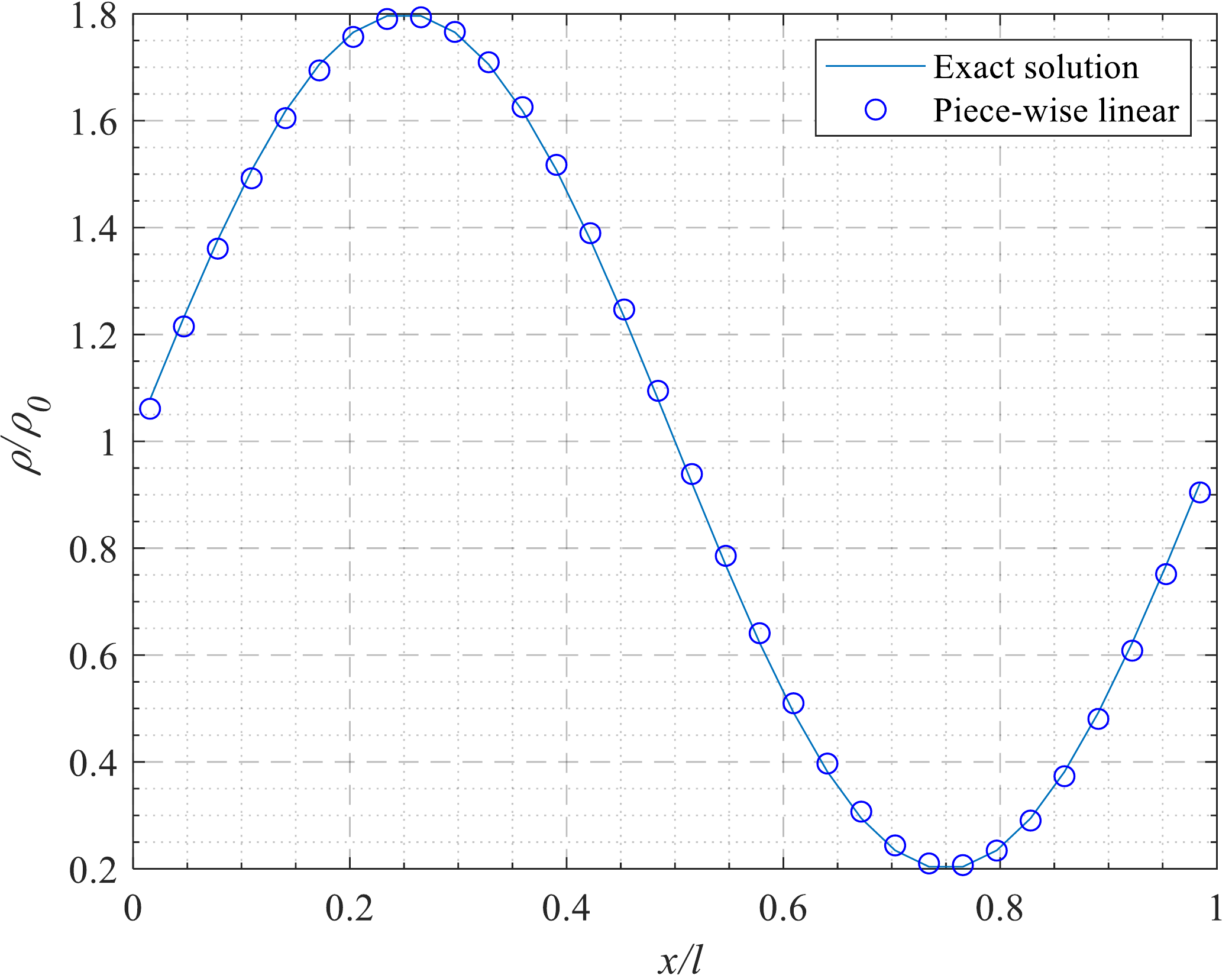}
    \caption{Comparison of exact solution with the numerical piece-wise linear solution obtained using the extended double-flux scheme with 32 cells.}
    \label{fig4-1-1}
\end{figure}

We use both the traditional conservative scheme and the extended double-flux scheme to calculate this problem, without any use of discontinuity limiters.
Unless otherwise specified, all subsequent extended double-flux method verification and validation results have been attained utilizing $\textit{Approach A}$. $\textit{Approach B}$ gives virtually identical results for this problem.
Figure \ref{fig4-1-1} illustrates that the numerical result obtained with the new extended double-flux scheme is consistent with the exact solution. As listed in Table \ref{tab1}, our new method even shows superior performance over the traditional conservative scheme in the $L_{2}$ error of density. While not specifically shown, it also performs similarly in the $L_{1}$ and $L_{\infty}$ error norms as well as errors norms of other quantities.
These results show that although some conservation loss is caused inevitably by the double-flux model, numerical errors as well as the convergence order get indeed improved by curing the oscillation issue in the presence of consistent $\gamma$ variation.
This test verifies that the proposed extended double-flux scheme works well in this one-dimensional case. 

\begin{table*}[h]
    \centering
    \renewcommand\arraystretch{1.2}
    \caption{Convergence study of the extended double-flux method for one-dimensional entropy wave advection.}
    \footnotesize
    \begin{tabular}{ccccc}
    \hline
          & \multicolumn{2}{c}{traditional ccheme} & \multicolumn{2}{c}{extended double-flux method} \\ \hline
    $h$     & $L_{2}$ error             & convergece order& $L_{2}$ error                 & convergece order\\
    1/16  & 0.05032             &   -               & 0.04854                  &      -              \\
    1/32  & 0.01263             & 1.9940           & 0.01182                  & 2.0374               \\
    1/64  & 0.003159            & 1.9991           & 0.002927                 & 2.0142               \\
    1/128 & 0.0007903           & 1.9993           & 0.0007301                & 2.0032               \\
    1/256 & 0.0001977           & 1.9991           & 0.0001823                & 2.0013               \\ \hline
    \end{tabular}
    \label{tab1}
\end{table*}

\begin{figure}[H]
    \centering
    \includegraphics[width=0.7\textwidth]{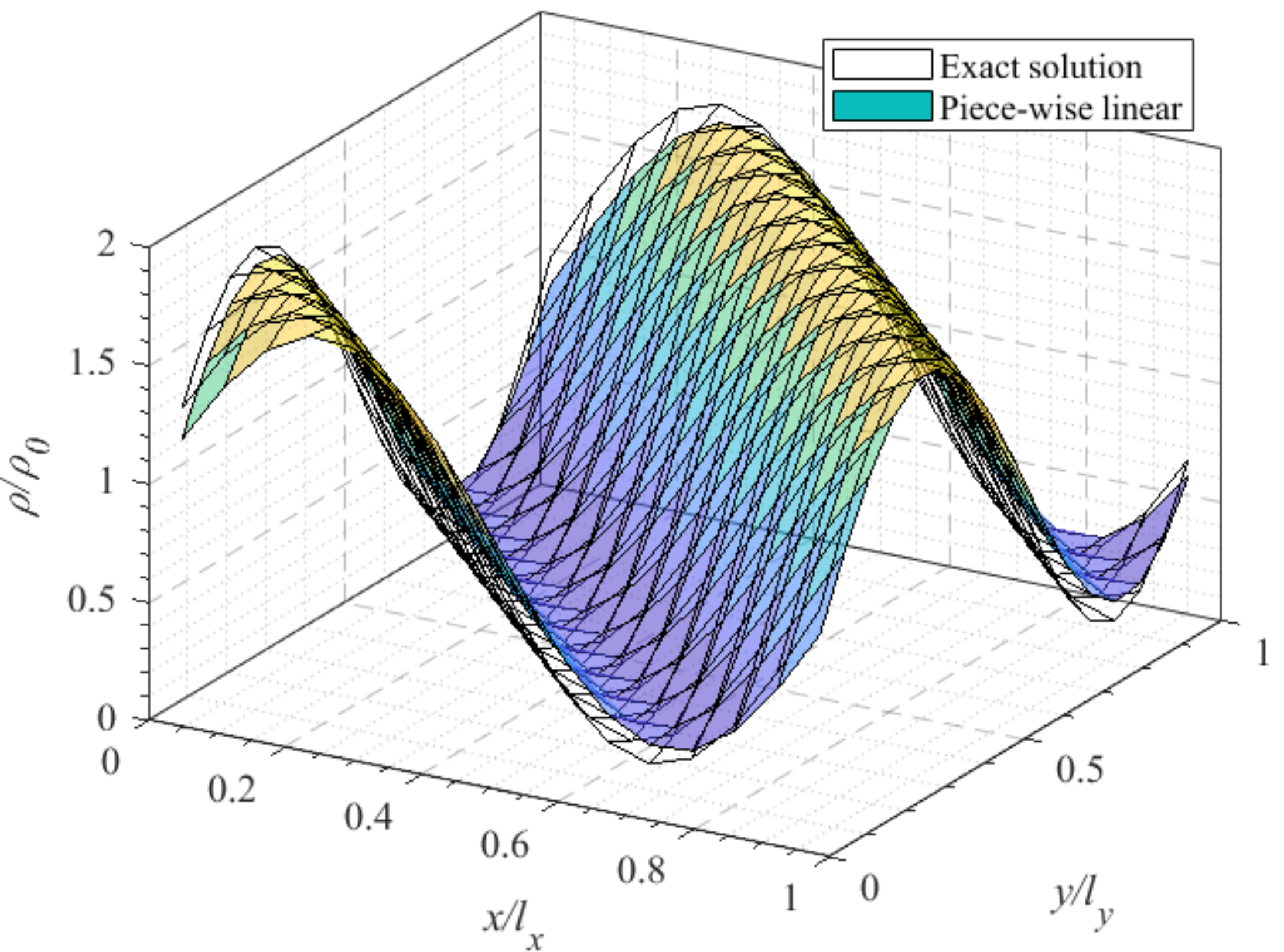}
    \caption{Comparison of exact solution with the numerical one obtained using the extended double-flux scheme with 16$\times$16 cells.}
    \label{fig4-1-2}
\end{figure}

To further verify our scheme in handling multi-dimensional smooth problems, we then extend this problem to two dimensions with the exact solution
\begin{equation}
\begin{array}{c}
\rho = h(x-u t-v t)+\rho_{\infty}, \\
u = u_{\infty}, \\
v = v_{\infty}, \\
p = p_{\infty}
\end{array}
\end{equation}
and periodic domain $\left(x,y\right) \in [0,1]\times [0,1]$.
Figure \ref{fig4-1-2} shows that the numerical solution coincides with the exact solution.
A convergence study is conducted to verify the global accuracy of the quasi-conservative scheme in handling multi-dimensional smooth problems. 
Results in Table \ref{tab2} illustrate that the quasi-conservative solution procedure performed well with second order accuracy in and with slightly higher convergence order when dealing with smooth variations of $\gamma$.

\begin{table*}[h]
    \centering
    \renewcommand\arraystretch{1.2}
    \caption{Convergence study of the extended double-flux method for two-dimensional entropy wave advection.}
    \footnotesize
    \begin{tabular}{ccccc}
    \hline
          & \multicolumn{2}{c}{traditional scheme} & \multicolumn{2}{c}{extended double-flux method} \\ \hline
    $h$     & $L_{2}$ error             & convergence order& $L_{2}$ error                 & convergence order\\
    1/16  & 0.09896             &   -               & 0.09537                  &      -              \\
    1/32  & 0.02533             & 1.9657           & 0.02369                  & 2.0088               \\
    1/64  & 0.006354            & 1.9953           & 0.005882                 & 2.0103               \\
    1/128 & 0.001591           & 1.9977           & 0.001466                & 2.0036               \\
    1/256 & 0.0003981           & 1.9988           & 0.0003661                & 2.0020               \\ \hline
    \end{tabular}
    \label{tab2}
\end{table*}

To evaluate the efficiency of the proposed method, we compare the absolute and relative wall cpu time of our presented solver and the original one in AMROC \cite{deiterding2003parallel}, as listed in Table \ref{tab_cutin}.
It can be observed that although implementing several additional steps like double-flux calculation and energy term modification on the traditional discretization procedure incurs some expenses, the relative cost is always below 10\% with the use of from 16 to 512 cells, which is quite acceptable.
Moreover, it should be noted that during the calculation of this special case, every cells employed the presented scheme during its discrete evolution while in a real-world computation only a partial region of the whole domain, i.e., the location of the material interfaces or shear layers, should adopt this correction procedure by leveraging the hybrid procedure that will be introduced in Section \ref{sec3.2} and thus the global computational cost increase would be marginal. 

\begin{table*}[]
    \centering
    \caption{Absolute and relative cpu wall-time consumption of the multi-dimensional entropy waves problem calculated by the fully conservative and extended double-flux schemes. The mesh resolution ranges from 1/16 to 1/512.}
    \footnotesize
    \begin{tabular}{cccllll}
    \hline
      & 1/16 & 1/32 & 1/64 & 1/128 & 1/256 & 1/512 \\ \hline
    \begin{tabular}[c]{@{}c@{}}$t_{\text{cpu, fc}}$ [s]\end{tabular} & 0.517 & 1.423 & 4.191 & 18.527 & 98.924 & 649.814 \\
    \begin{tabular}[c]{@{}c@{}}$t_{\text{cpu, df}}$ [s]\end{tabular} & 0.521 & 1.495 & 4.413 & 19.147 & 106.954 & 697.568 \\
    \begin{tabular}[c]{@{}c@{}}$\frac{t_{\text{cpu, df}}-t_{\text{cpu, fc}}}{t_{\text{cpu, fc}}}$ [\%]\end{tabular} & \multicolumn{1}{l}{\color{red}0.774} & \multicolumn{1}{l}{\color{red}5.060} & \color{red}5.297& \color{red}3.346& \color{red}8.117& \color{red}7.3490\\ \hline
    \end{tabular}
    \label{tab_cutin}
\end{table*}

\subsubsection{1-D advection of an inert \ce{H2} bubble in \ce{O2} gases}

To verify the proposed two new approaches, as well as comparing their abilities to resolve the contact surfaces, the advection of an $\ce{H2}$ bubble in $\ce{O2}$ \cite{bando2020comparison} is studied here by solving the inviscid Euler equations with the numerical scheme discussed in Section \ref{sec2.2}.
This example illustrates the classic oscillation phenomena when encountering a discontinuous $\gamma$ with a fully conservative scheme, as well as how this is overcome by our new extended double-flux method.

The $\ce{H2}$ bubble is advected with a constant velocity $u_{0}$ = 20 m/s and a uniform pressure $p_{0}$ = $1 \times 10^{5}$ Pa. 
The periodic boundary condition is applied on both left and right boundaries.
The initial profiles of mass fraction $Y_{\ce{H2}}$ and temperature $T$ are defined as
\begin{subequations}
\begin{align}
    T(x)&=\frac{1}{2}[(1+\Theta)+(1-\Theta)(\tanh (|x-x_{0}|-d)), \\
    Y_{\mathrm{H}_{2}}(x)&=\frac{1}{2}[1-\tanh (|x-x_{0}|-d)],
\end{align}
\end{subequations}
where the centre of the plateau $x_{0}$, the width of the plateau $d$ and the temperature ratio $\Theta$ are set to 0, 10 and 7, respectively and as in \cite{bando2020comparison}.
The physical length of the domain is 50 cm with a uniform grid size $\Delta x$ = 0.2 cm.
This test is run with a CFL number of 0.4.
The $\ce{H2}$ bubble propagates within the computational domain before the end of the calculation $t_{e}=8500\delta t$ ($\delta t\approx2.355\times10^{-7}s$).

\begin{figure}[H]
\centering
    \centering
    \begin{subfigure}{1.0\textwidth}
        \centering
        \includegraphics[width=0.65\textwidth]{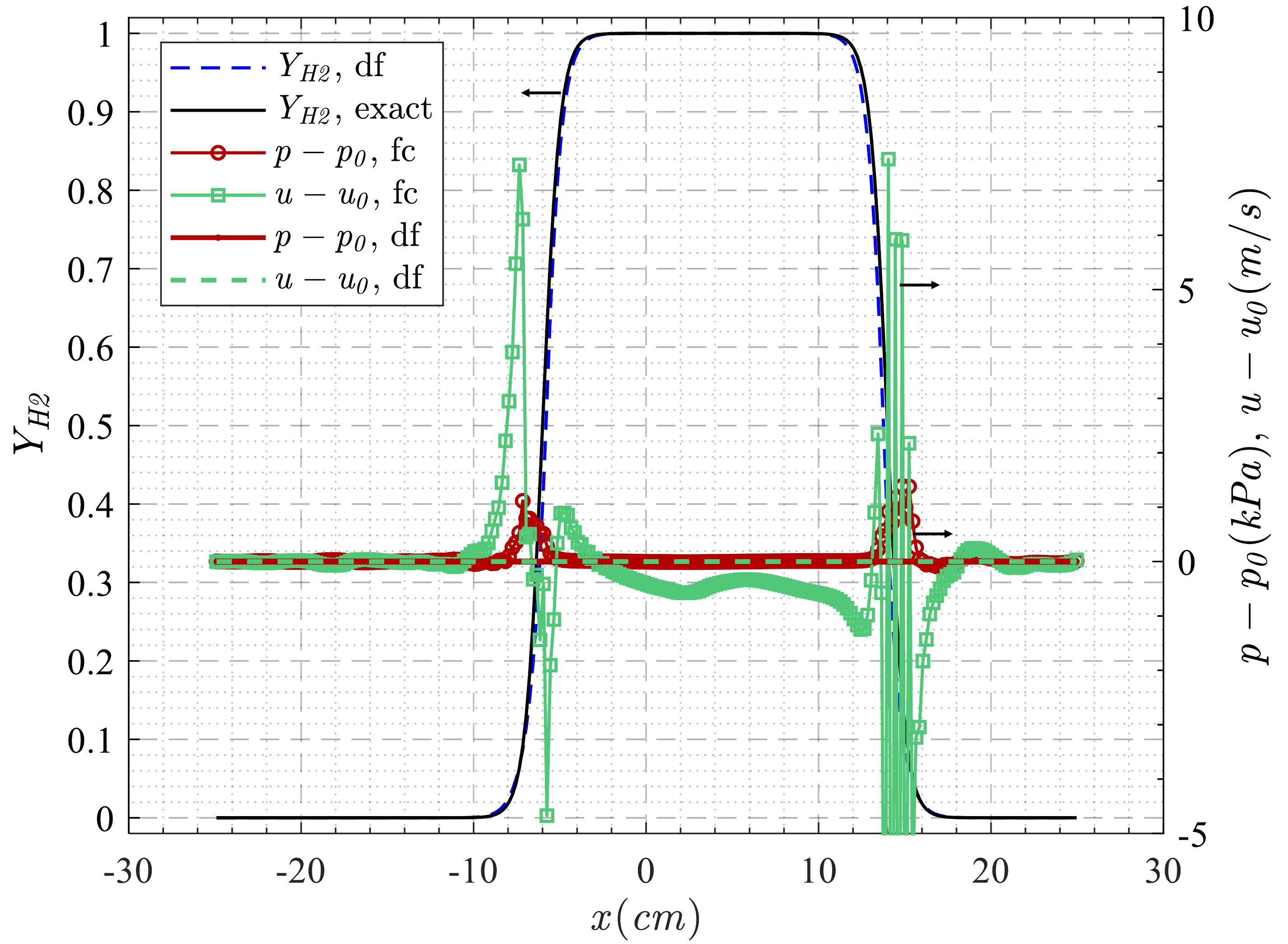}
        \caption{}
        \label{fig4-2-1(a)}    
    \end{subfigure}
    \begin{subfigure}{1.0\textwidth}
        \centering
        \includegraphics[width=0.65\textwidth]{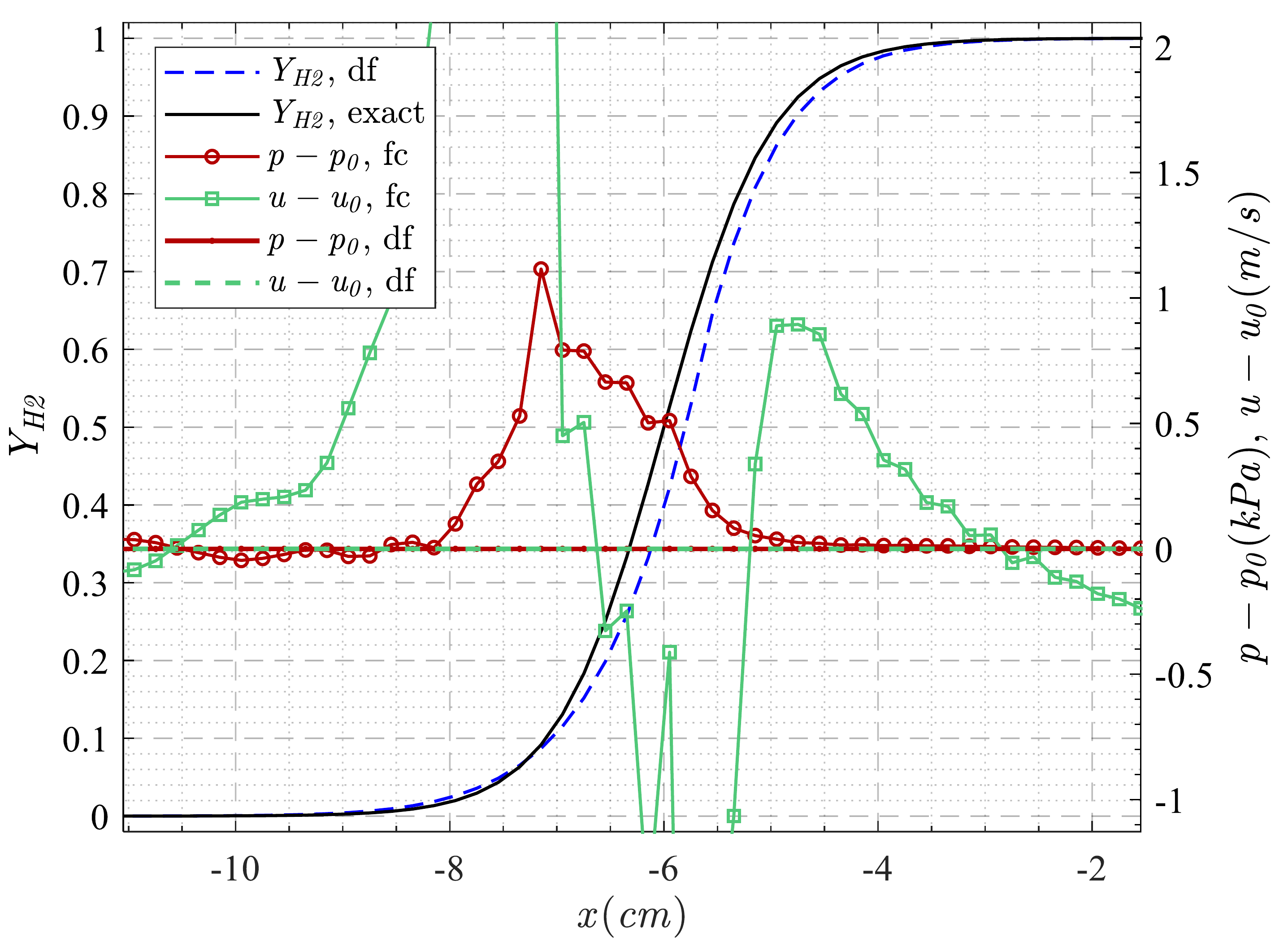}
        \caption{}
        \label{fig4-2-1(b)}    
    \end{subfigure}   
\caption{(a) 1-D inert $\ce{H2}$ bubble advection solved by the extended double-flux solver (df) and the fully conservative scheme (fc) at $t=t_{e}$, (b) local enlargement of Fig. \ref{fig4-2-1(a)}.}
\label{fig4-2-1}
\end{figure}

The numerical results from the standard conservative scheme and our new presented approaches are depicted in Fig. \ref{fig4-2-1}.
For clarity, we only plot the results obtained with one double-flux approach, since the other performed virtually identically. 
According to these results, the symmetry of the $\ce{H2}$ bubble as well as the pressure equilibrium are both perfectly preserved by the extended double-flux approach, as seen in the local enlargement of Fig. \ref{fig4-2-1(b)}. On the other hand, the fully conservative scheme produces pressure and velocity oscillations clearly visible in Figs. \ref{fig4-2-1(a)} and \ref{fig4-2-1(b)}.

\begin{figure}[H]
\centering
    \centering
    \begin{subfigure}{1.0\textwidth}
        \centering
        \includegraphics[width=0.65\textwidth]{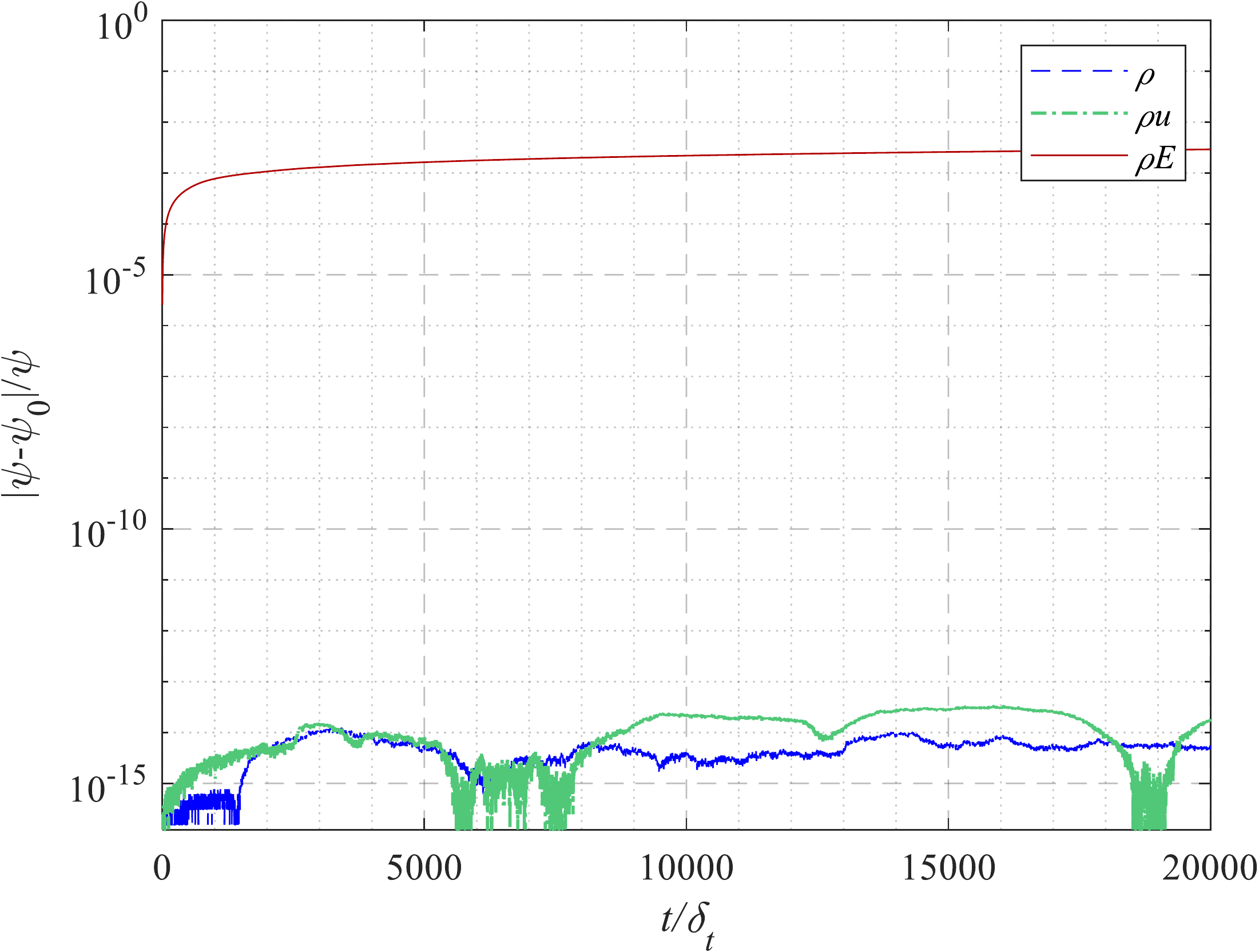}
        \caption{}
        \label{fig4-2-2(a)}    
    \end{subfigure}
    \begin{subfigure}{1.0\textwidth}
        \centering
        \includegraphics[width=0.65\textwidth]{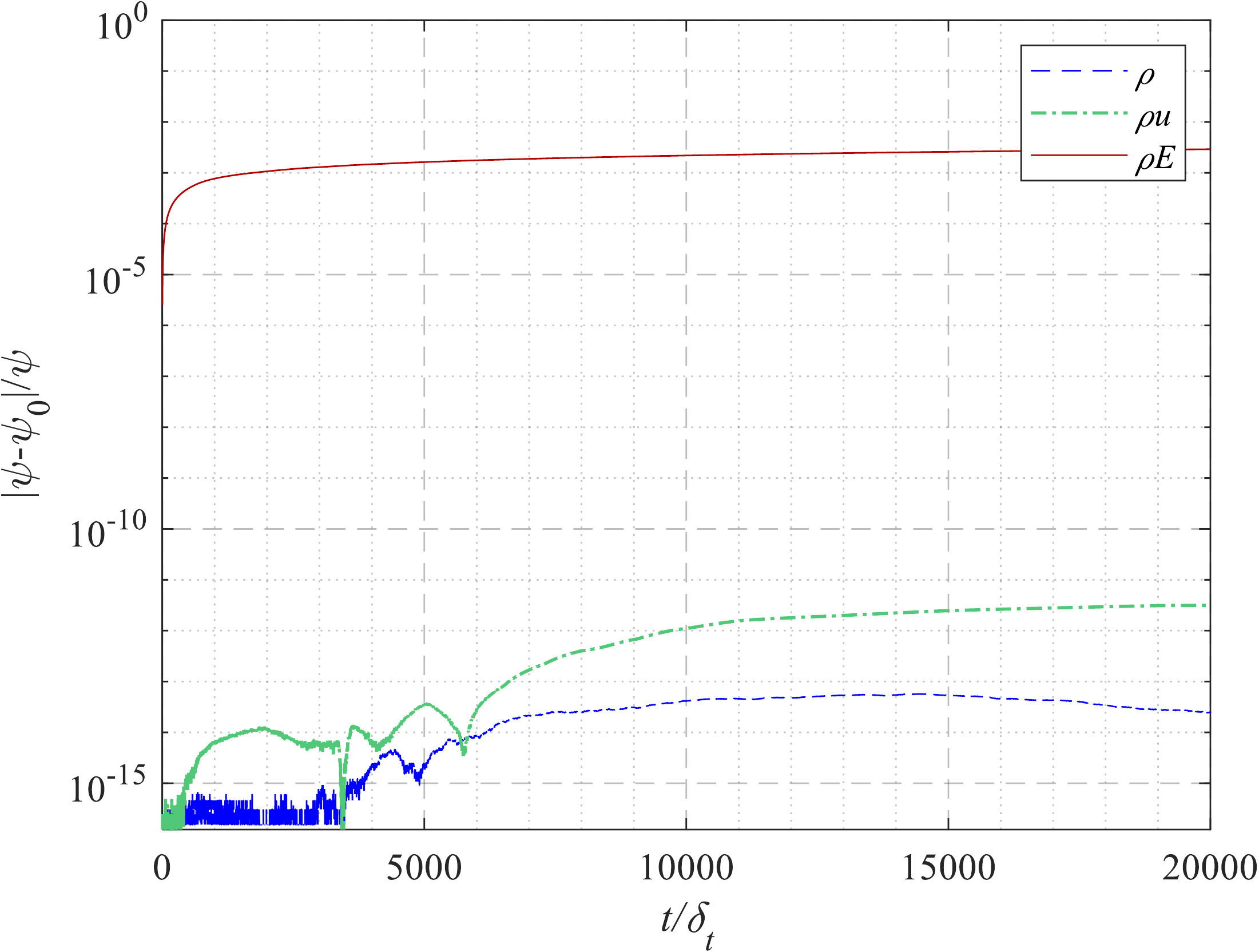}
        \caption{}
        \label{fig4-2-2(b)}    
    \end{subfigure}   
\caption{Conservation loss by the extended double-flux method with (a) $\textit{Approach A}$ and (b) $\textit{Approach B}$.}
\label{fig4-2-2}
\end{figure}

The conservation loss is recorded during the time span $\left[0,t_{e}\right]$, with the error at time step $n$ expressed as 
\begin{equation}
    e^{n} = \frac{\left|\psi^{n}-\psi_{0}\right|}{\psi^{n}}= \frac{\left|\Sigma_{j=1}^{N}\left(U^{n}_{j}-U^{0}_{j}\right)\right|}{\Sigma_{j=1}^{N}U^{n}_{j}}
\end{equation}
with $U_{j}^{n}=\left[\rho,\rho u,\rho E\right]^{n}$. Both extended double-flux approaches produce a conservation error in mass and momentum within the machine accuracy, as seen in Fig. \ref{fig4-2-2(a)} and \ref{fig4-2-2(b)}.
Note that the final total energy loss in this problem is with 0.002913 also significantly smaller as the results in \cite{houim2011low}, which is achieved thanks to the use of one auxillary variable and the proper reformulation of $E$. 
For mass and momentum conservation, we note that $\textit{Approach A}$ gives slightly better performance than $\textit{Approach B}$ in momentum and with comparable performance in mass conservation.
Both approaches have been confirmed to capture material interfaces with no artificial oscillations.

\section{An efficient, adaptive solver for simulation of shock-interface interaction}
\label{sec3}
In spite of our efforts, the inherent conservation concerns accompanied with the extended double-flux method remain an issue. Hence, we have designed an explicit hybrid methodology that ensures correct shock capturing nevertheless. In addition, we have incorporated a highly efficient adaptive mesh refinement (AMR) technique into the overall solver.
\subsection{Hybridization methodology}
\label{sec3.2}
In compressible flows, shock waves are ubiquitous, e.g., in shock bubble interaction, shock induced combustion or detonation.
In that respect, a fully conservative shock capturing method is usually preferred under those circumstances to make sure one obtains the correct weak solution.
Using a non-conservative scheme like the double-flux method, on the other hand, is generally not a good choice because it might generate huge numerical artifacts by producing an unphysical shock.
To remedy this deficiency,  we apply hybridized numerical fluxes in the present study by combining the fully conservative shock-capturing scheme and the quasi-conservative extended double-flux method.

To achieve this, we utilize the normalised pressure curvature \cite{hill2004hybrid} as a shock sensor
\begin{align}
S_{i} & = \frac{\left|\bar{p}_{i-1}-2 \bar{p}_{i}+\bar{p}_{i+1}\right|}{\bar{p}_{i-1}+2 \bar{p}_{i}+\bar{p}_{i+1}}.
\end{align}
This normalized sensor is generally resolution independent and hence directly compatible with the local grid refinement of the AMR technique employed. The working procedure is as follows and summarized in Algorithm \ref{al1}.
Before time marching each computational cell, the sensor value $S_{i}$ is calculated to determine which scheme (conservative or non-conservative) is selected. 
Once $S_{i}$ exceeds a prescribed threshold $S_{i}^{t}$, the double-flux solver will reduce back to the traditional conservative solver by jumping over the frozen and flux correction procedures.
If not, a quasi-conservative scheme (aka the extended double-flux ) would be applied on this cell with the use of either $\textit{Approach A}$ or $\textit{B}$.
After that, a correction procedure is implemented to update the thermodynamic quantities as well as the total energy $\rho E$.
Determination of the threshold $S_{i}^{t}$ is naturally problem-related, which is a common issue of hybrid methods, cf. \cite{ziegler2011adaptive}. 
In most occasions, they can be resolved separately with the use of our presented hybrid solver.
For phenomena like the detonation wave in whose structure the shock and reaction fronts are coupled closely, it should be and also is resolved simply by the conservative scheme since this phenomenon is dominated more by shock compression instead of physical diffusion and thereby ensuring the correct weak solution is critical.

The extended double-flux scheme is a Godunov-type scheme and is usually extended to higher order accuracy in space and time \cite{billet2003adaptive}. 
For the MUSCL-TVD scheme employed in this work, we interpolate the primitive variables onto the cell edges, which is computationally efficient and preferred in most occasions.
However, as reported in \cite{houim2011low}, this interpolation technique may generate oscillations near shock fronts.
Although this can be tackled to some degree by decomposing the interpolation variables onto the characteristic fields globally, the entire computational load would be tremendously increased and also the double-flux algorithms would become much more involved \cite{houim2011low}.
Consequently, in this paper, we also combine the selection of interpolation variables into our current hybrid solver: we decompose the conservative system into its characteristic form as in \ref{apB} in the regions of shocks but still use primitive variables in smooth regions within the framework of the extended double-flux scheme.
This avoids complicating the the double-flux algorithms unnecessarily as well as guaranteeing the conservation and stability at the shocks.
For the compared standard conservative scheme, we just use primitive variable interpolation globally.
This hybridization is extensively validated in Secs. \ref{sec4.4}, \ref{sec4.5}, \ref{sec4.7} and \ref{sec4.8}.

\subsection{Adaptive mesh refinement}
\label{sec3.3}
To handle realistic compressible flow and combustion problems with high Reynolds number, it is beneficial to supplement the new hybrid scheme with dynamic mesh adaptivity.  
Therefore, to suffiently resolve the small-scale flow structures in large-scale domains, a blockstructured adaptive mesh refinement (SAMR) technique pioneered by Berger and Oliger \cite{berger1989local} is employed in this work.
In the present paper, all the SAMR procedures are carried out by the object-oriented AMROC (Adaptive Mesh Refinement in Object-oriented C++) framework \cite{deiterding2011block}.
AMROC allows a fully autonomous rewriting of the grid-wise numerical update in Fortran whereas the parallel AMR is organized independently as the outer framework in C++.
It should be noted that the impacts of grid-level operations, such as the coarse-fine prolongation and restriction, on the non-conservative double-flux scheme have not been taken into account and hence here we have always ensured that the material interface structure is properly nested within the highest refinement level.

Since the AMROC software has been extensively validated for its AMR algorithm in previous literature \cite{deiterding2003parallel,deiterding2009parallel,cai2018experimental}, we will not discuss the verification of AMR procedures in this paper.

\subsection{Verification and validation}
\label{sec4.2m}
\subsubsection{1-D Sod problem}
\label{sec4.4}
This standard benchmark tests the shock-interface capturing ability of our presented hybrid scheme. 
A typical one-dimensional shock tube problem with single component is calculated.
The initial condition for this example is
\begin{equation}
  \left(\ T\text{(K)},\ u\text{(m/s)},\ p\text{(MPa)},\ Y_{\ce{N2}}\ \right)=    \left\{
    \begin{aligned}
        & \left(300,\ 0,\ 0.5,\ 1.0\right)_\textit{L} \\
        & \left(2000,\ 0,\ 0.1,\ 1.0)\right)_\textit{R}
    \end{aligned},
    \right.
\end{equation}
with the initial discontinuity placed at $x=5$ cm and the entire length of the domain was chosen as 15 cm.

\begin{figure}[H]
    \centering
    \includegraphics[width=0.65\textwidth]{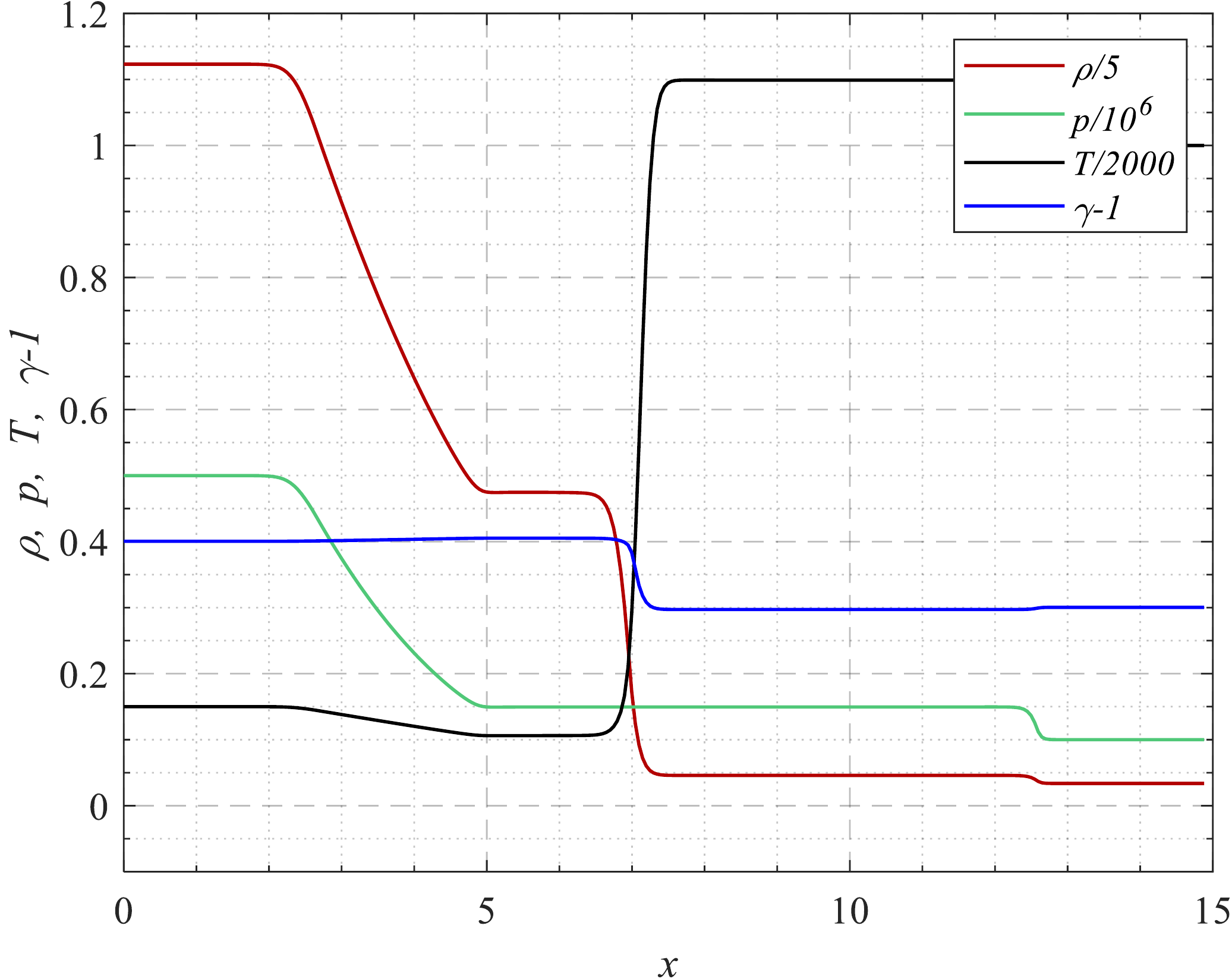}
    \caption{Quantity profiles of 1-D Sod problem solved by the hybrid fluxes.}
    \label{fig4-4-1}
\end{figure}

\begin{figure}[H]
    \centering
    \includegraphics[width=0.65\textwidth]{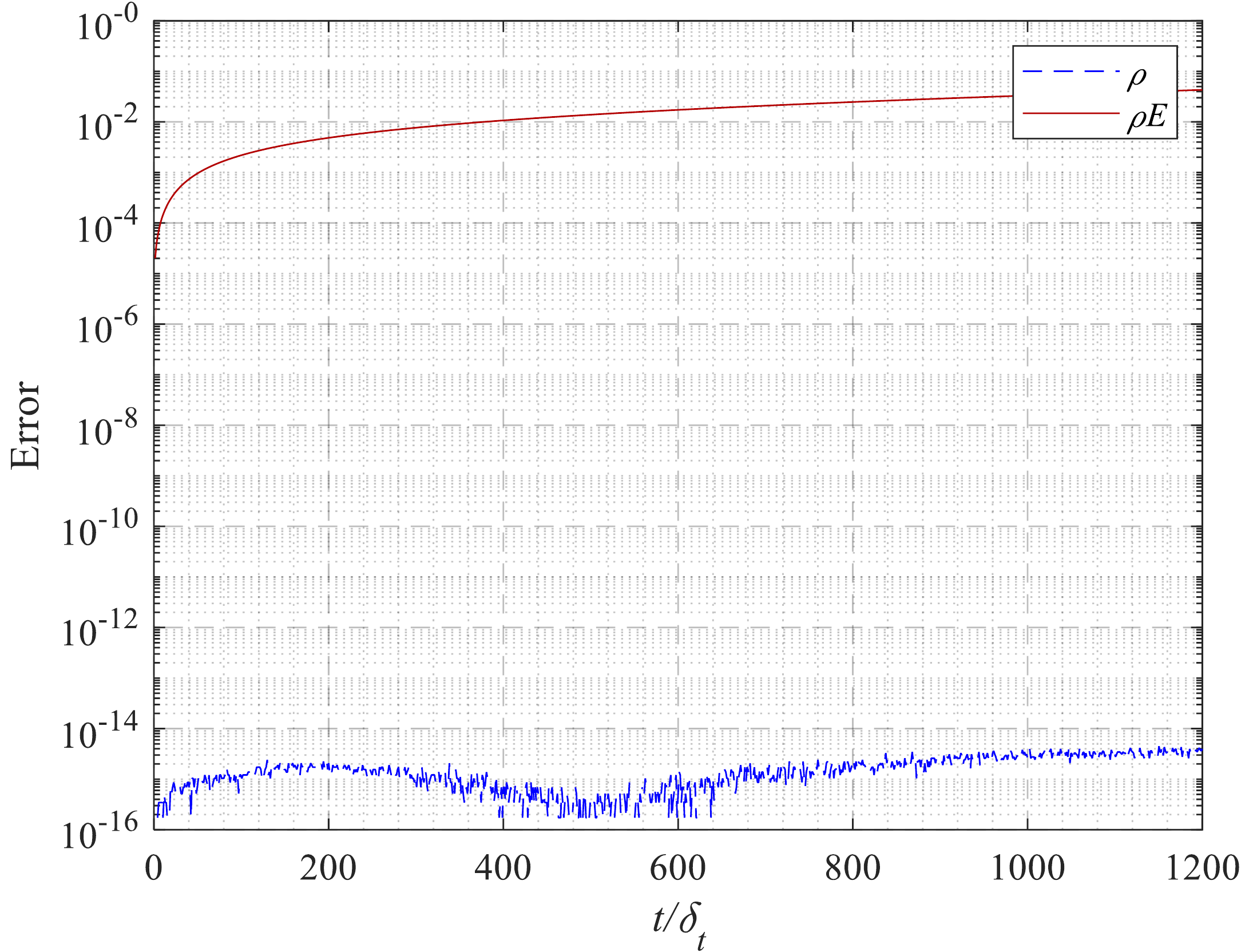}
    \caption{Conservation loss of mass and total energy in 1-D Sod problem solved by hybrid fluxes.}
    \label{fig4-4-2}
\end{figure}

\begin{figure}[H]
\centering
    \centering
    \begin{subfigure}{1.0\textwidth}
        \centering
        \includegraphics[width=0.65\textwidth]{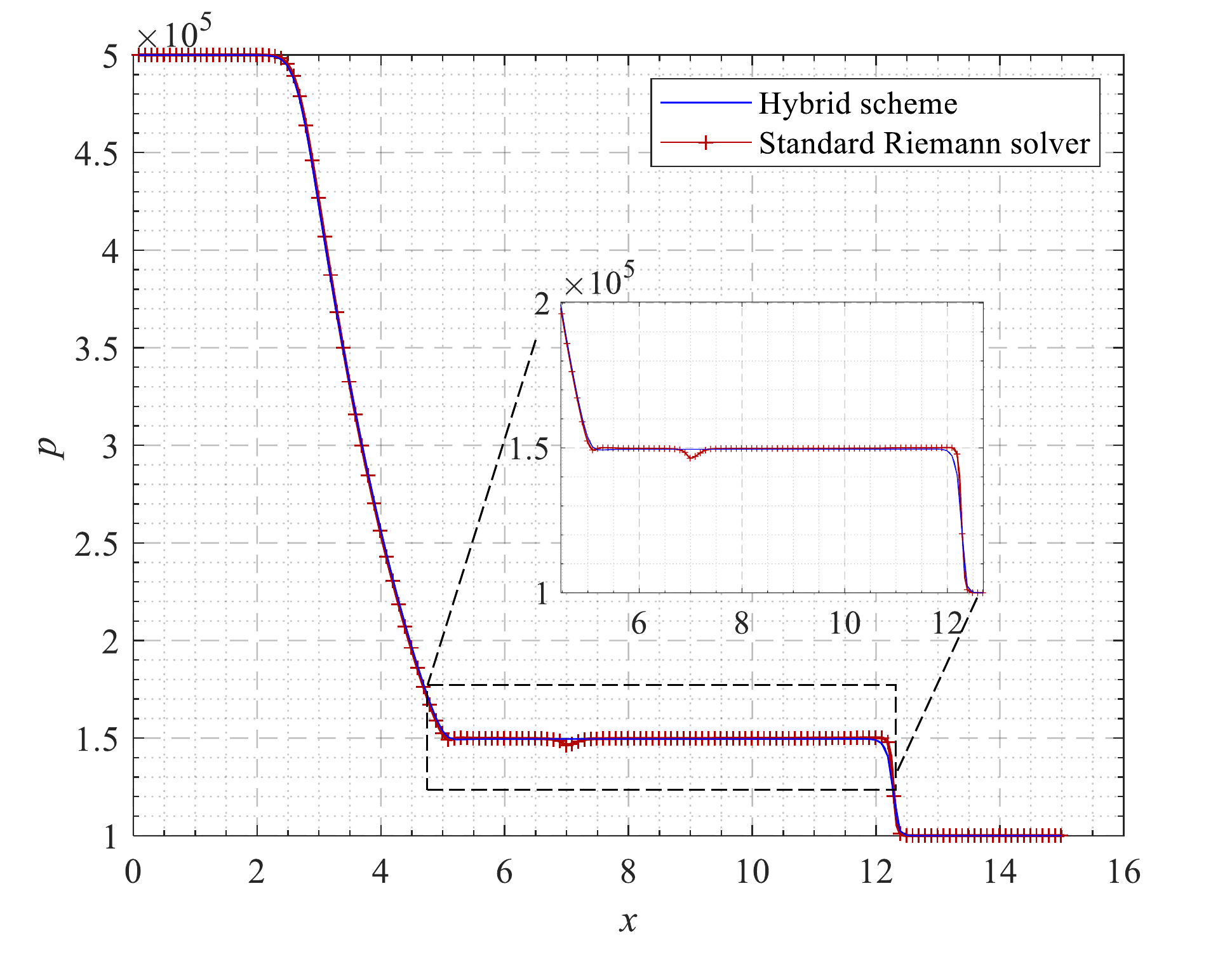}
        \caption{}
        \label{fig4-4-3}    
    \end{subfigure}
    \begin{subfigure}{1.0\textwidth}
        \centering
        \includegraphics[width=0.58\textwidth]{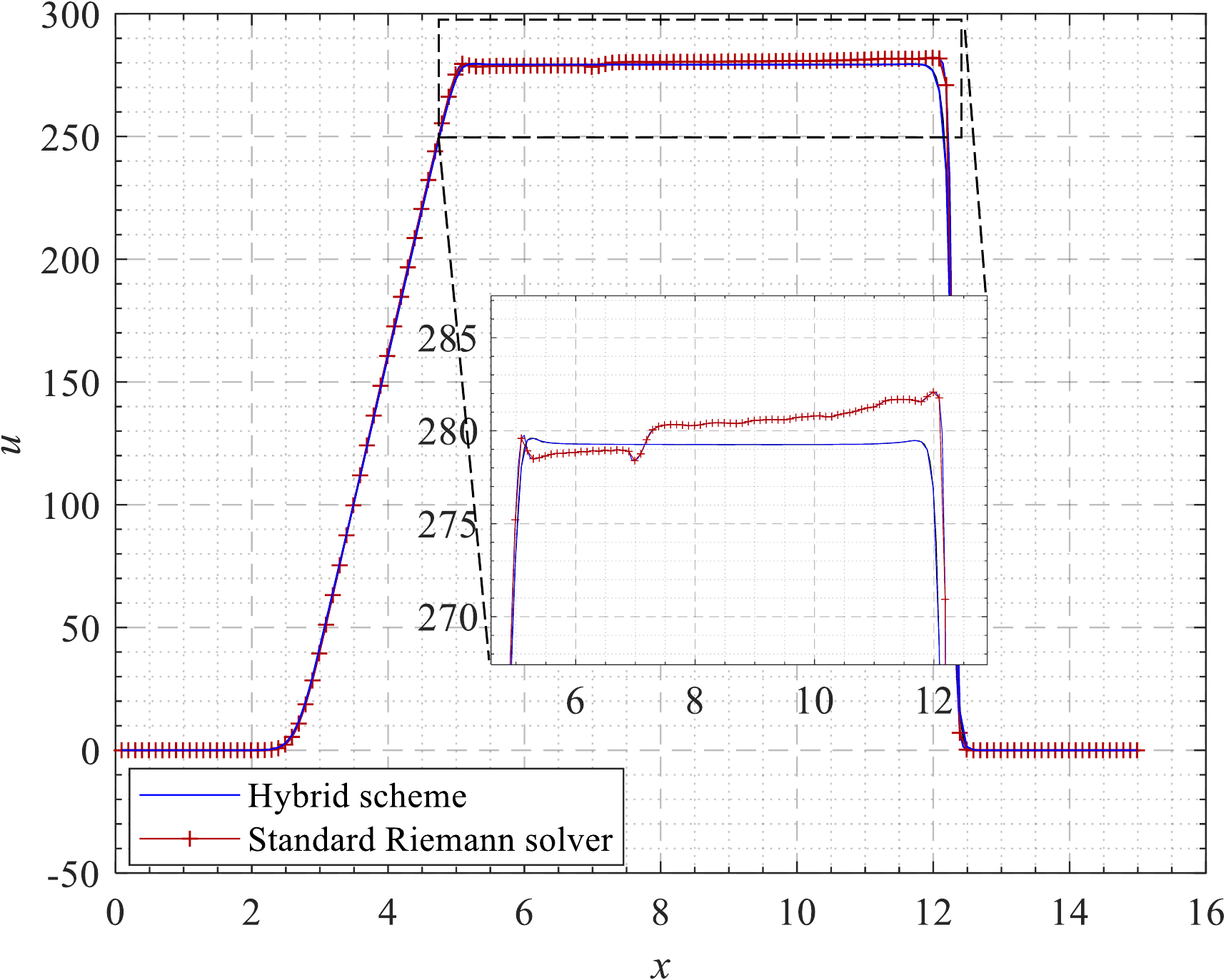}
        \caption{}
        \label{fig4-4-4}    
    \end{subfigure}   
\caption{Comparison of the 1-D (a) pressure and (b) velocity profiles of the 1-D Sod problem solved by the hybrid scheme and the conservative Riemann solver.}
\end{figure}

In the hybrid solver as introduced in Algorithm \ref{al1}, the primitive interpolation as well as the extended double-flux method is applied at the material interfaces, while the non-linear waves, i.e., the shock wave and rarefaction waves, are captured via the fully conservative scheme with a more stable but inefficient characteristic decomposition procedure. 
The numerical solution and the conservation loss at $t=100\mu$s are shown in Figs. \ref{fig4-4-1} and \ref{fig4-4-2}.
Considering the difficulty of obtaining an exact solution of the Sod problem for thermally perfect mixtures whose specific heat ratio $\gamma$ varies with the temperature, we compare here the results obtained with the traditional conservative scheme as well as the hybrid scheme as in Figs. \ref{fig4-4-3} and \ref{fig4-4-4}.

Compared to \cite{houim2011low}, the conservation loss of $\rho$ and $\rho E$ is significantly reduced due to the proper treatments to the non-linear waves. 
In \cite{houim2011low}, the mass error reached about $10^{-3}$ while here it approaches roughly the round-off error.
It must be noted that the energy loss happens only at the material front instead of the shock. 
Figures. \ref{fig4-4-3} and \ref{fig4-4-4} confirm that the hybrid solver converges to the same weak solution correctly as compared to the fully conservative one.  
In Fig. \ref{fig4-4-3}, it can be observed that an obvious pressure spike emerging at the material front solved by the traditional conservative scheme is perfectly eliminated by our developed scheme. 
One can also find in Fig. \ref{fig4-4-4} that the velocity oscillation at the material front as well as in the post-shock regions is greatly suppressed due to the hybrid fluxes and interpolation techniques.
This example demonstrates the developed hybrid solver in handling one-dimensional shock-interface problems stably and accurately.


\subsubsection{2-D inert shock-bubble interaction}
\label{sec4.5}
\begin{figure}[H]
    \centering
    \includegraphics[width=0.9\textwidth]{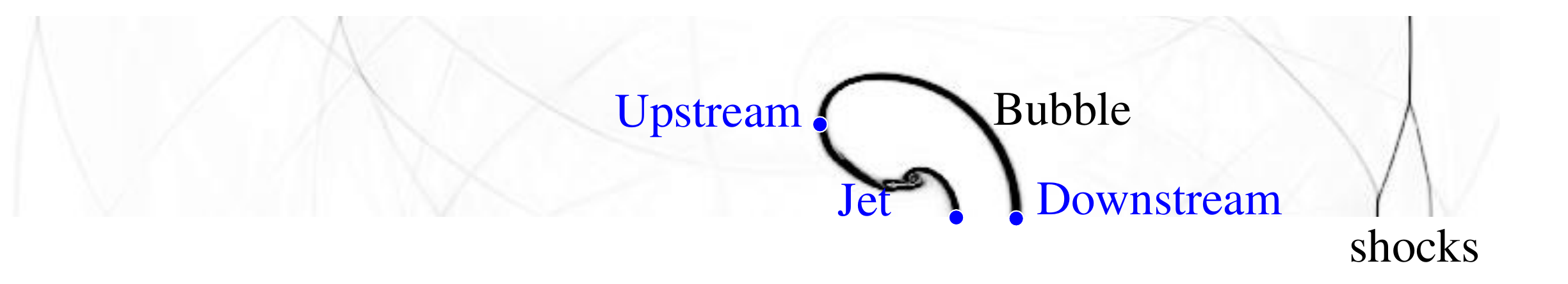}
    \caption{Scenario of a helium bubble deformed by a normal shock. Three time-dependent characteristic points are labelled respectively as the "upstream", "downstream" and "jet" points.}
    \label{fig4-5-0}
\end{figure}

To further valid the hybrid solver in handling multi-dimensional viscous problems, a 2-D inert shock-bubble interaction example was simulated by solving the viscous Navier-Stokes equations.
This example involves the passage of an air shock wave through a helium bubble suspended in the medium, as depicted in Fig. \ref{fig4-5-0}. 
In the past, this example has frequently served as a benchmark for two-dimensional simulation validations for interface-capturing methods \cite{quirk1996dynamics,terashima2009front,johnson2020conservative}.
The computational configuration is a rectangular domain defined by $\Omega = \left[0, 0.325\right]$ m $\times \left[0, 0.0455\right]$ m. 
The simulation starts with a normal shock wave situated at $\bar{x}$ = 0.225 m, preceded by the placement of the helium bubble on its left side.
A uniform grid of $\Delta x=\Delta y$ = 0.01 cm is employed.

\begin{figure}[H]
    \centering
    \includegraphics[width=0.8\textwidth]{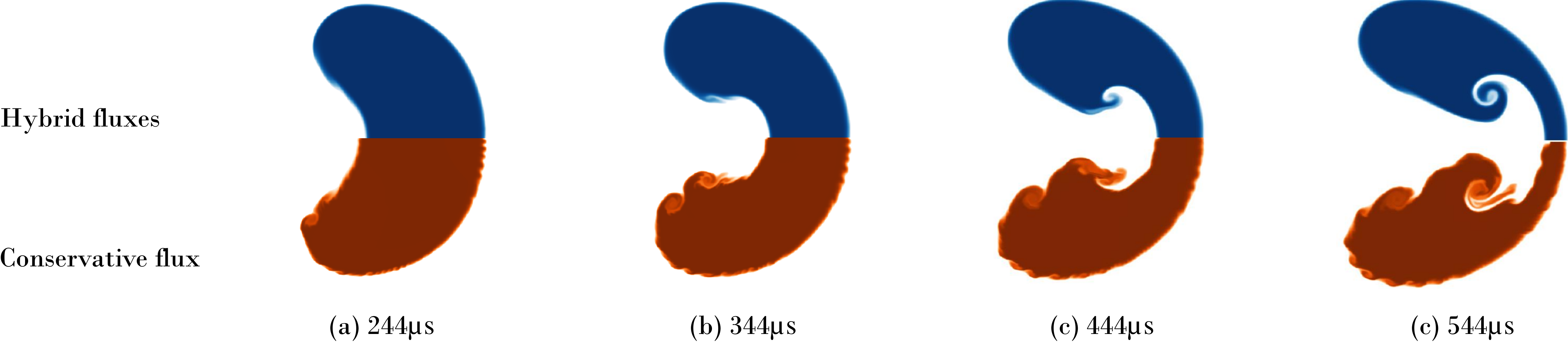}
    \caption{Comparison of the bubble deformation process calculated by the hybrid and the conservative schemes.}
    \label{fig4-5-1}
\end{figure}

\begin{figure}[H]
\centering
    \centering
    \begin{subfigure}{1.0\textwidth}
        \centering
        \includegraphics[width=0.55\textwidth]{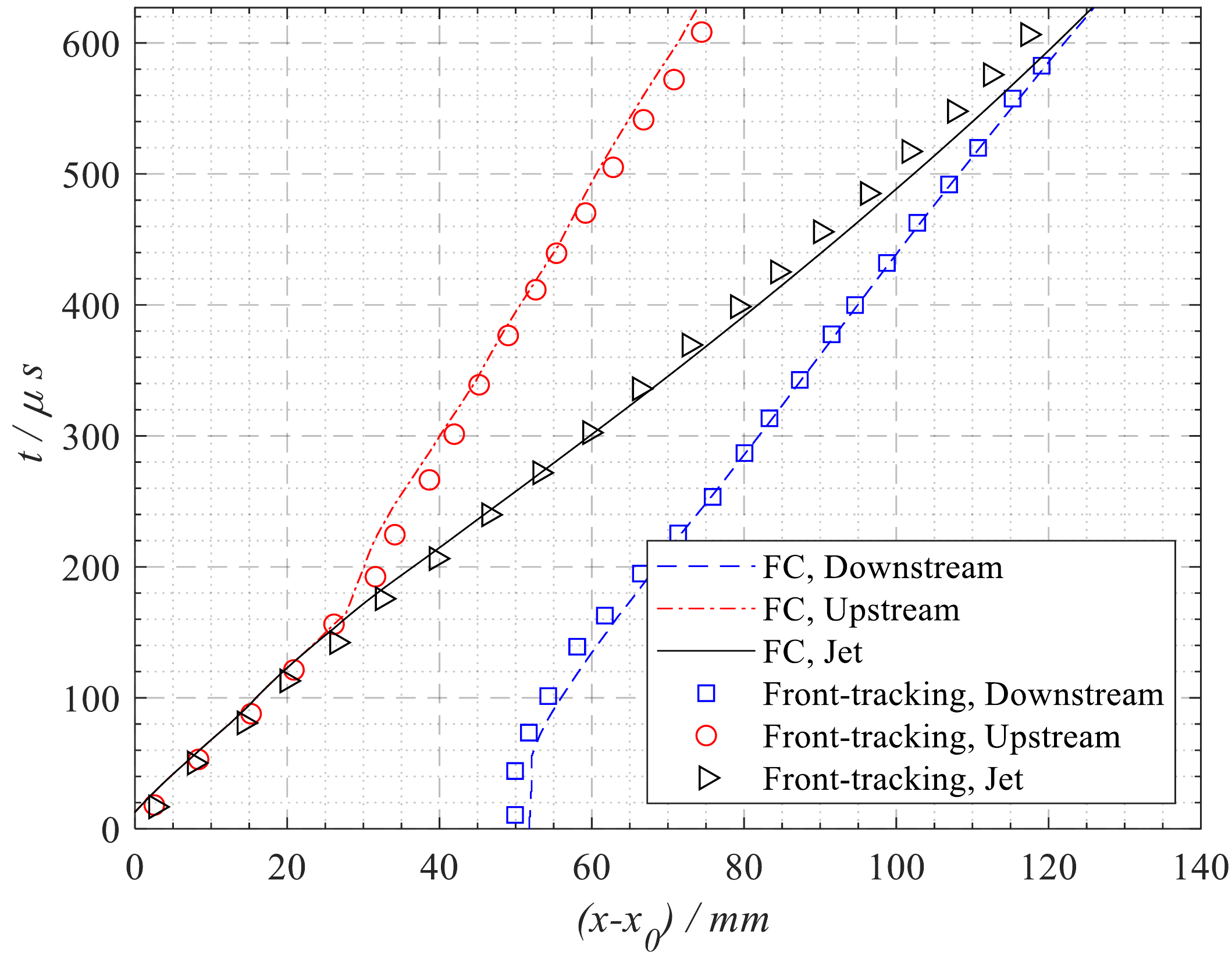}
        \caption{}
        \label{fig4-5-2}    
    \end{subfigure}
    \begin{subfigure}{1.0\textwidth}
        \centering
        \includegraphics[width=0.55\textwidth]{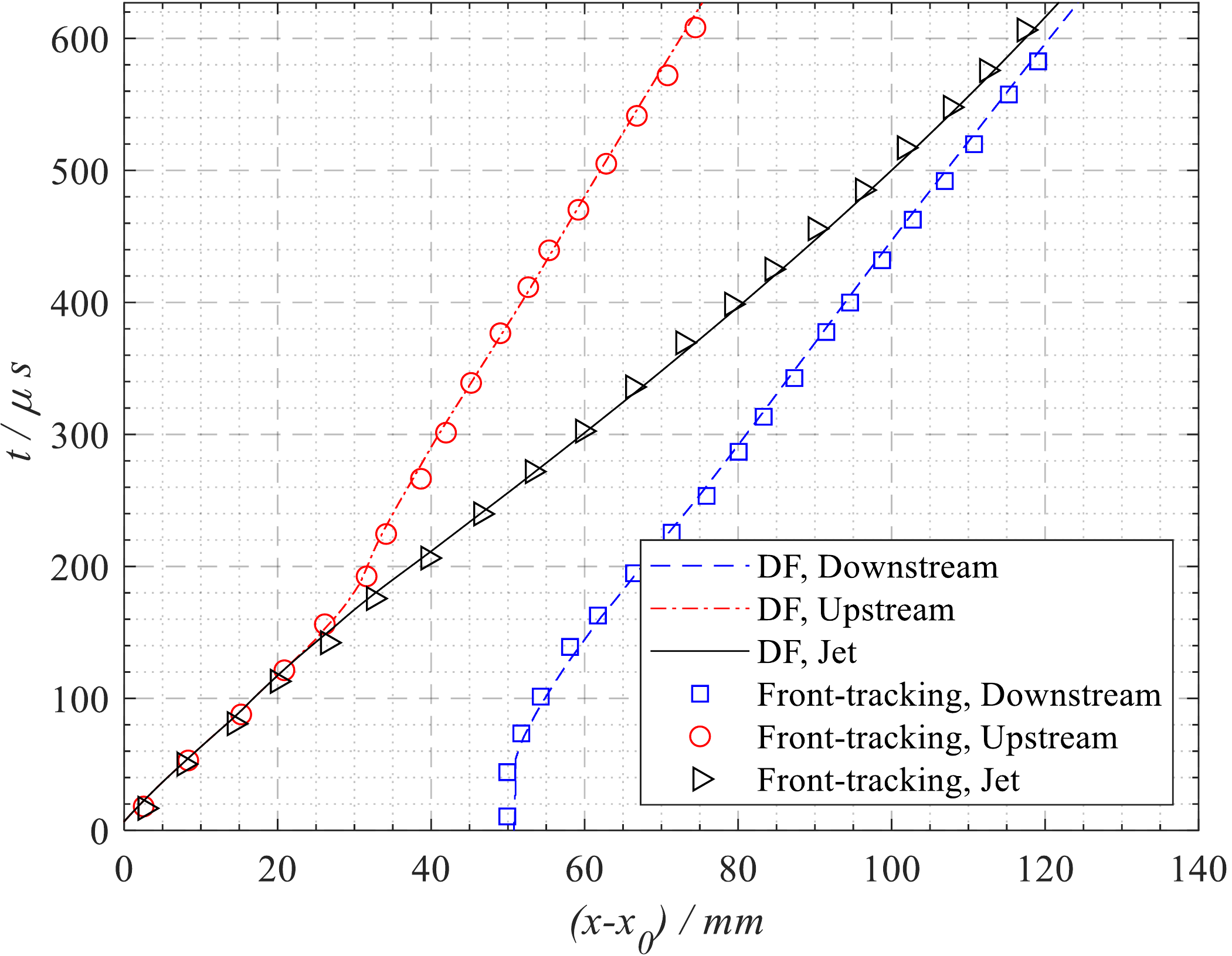}
        \caption{}
        \label{fig4-5-3}    
    \end{subfigure}   
\caption{Trajectory of three characteristic points obtained with the (a) conservative flux and (b) hybrid fluxes, as compared to the results calculated by the front-tracking method of \cite{terashima2009front}.}
\end{figure}

It can be observed in Fig. \ref{fig4-5-1} that the conventional conservative shock-capturing scheme introduces excessive spurious numerical oscillations when dealing with the interaction process between the shock and bubble, resulting in a visibly unstable structure of the helium plume, which should have been properly dissipated by the viscosity in the Navier-Stokes equations.
In contrast, the hybrid scheme yields a significantly smoother bubble interface and effectively exhibits the diffusion effect.
It is evident that the use of conservative scheme leads to a noticeable expansion in bubble volume over time, which can also be inferred from Fig. \ref{fig4-5-2}. 
Here, the dotted markers represent the bubble trajectories calculated using a front-tracking method \cite{terashima2009front}. 
Clear discrepancies emerge between the predictions of the front-tracking method and the traditional conservative scheme concerning the "upstream" and "jet" points, exacerbated over time due to numerical oscillation errors. 
This amplification of differences suggests a numerical inflation of the bubble volume, highlighting a failure in accurate numerical prediction. 
Conversely, Figure \ref{fig4-5-3} illustrates that results obtained with the improved hybrid approach align more favorably with the one from the sharp-interface capturing method.
This example also verifies that the multi-dimensional solution procedure of Algorithm \ref{al1} works reliably.

\section{Chemically reactive problems}
\label{sec4}
So far, we have constructed an efficient, accurate hybrid solver tailored especially for thermally perfect gases. 
It remains stable under shock capturing by the TVD limiting and the characteristic decomposition procedures and is capable of resolving material interfaces free of oscillations and with marginal conservation loss.
The AMR technique is also employed to achieve efficient computations.
Section \ref{sec3.2m} demonstrated the preferable properties of the extended double-flux scheme.
The hybridization methodology has been verified and validated in Section \ref{sec4.2m}, and its combination with AMR will be presented to be working well in the subsequent Sections \ref{sec4.7}. In the final section, an example of a complex planar RDE (Rotating Detonation Engine) with non-premixed gases will illustrate the superiority of the method developed in this paper. 

\subsection{1-D advection-diffusion of a reactive $\ce{H2}$ bubble in $\ce{O2}$ gases}
\label{sec4.3}

In this example, the complete multicomponent Navier-Stokes equations with diffusion and chemistry terms are solved.
We use this example to elucidate the failure of a fully conservative method as well as the capacity of the extended double-flux scheme in resolving a reactive contact discontinuity with minor pressure gradient. 

The initial and physical boundary conditions are set up following Ref. \cite{billet2003adaptive}, Section3.2.
The initial profiles of $Y_{\mathrm{H}_{2}}$ and $T$ are
\begin{subequations}
\begin{align}
T &= 1500\left[1-\frac{1}{3} \tanh \left(C_{r}\left(\frac{l}{2}-\left|x-x_{0}\right|\right)\right)\right],\\
Y_{\mathrm{H}_{2}} &= \frac{1}{2}\left[1+\tanh \left(C_{r}\left(\frac{l}{2}-\left|x-x_{0}\right|\right)\right)\right],\\
u_{0} &= 10 m/s, \quad \quad \quad p_{0} = 1\times10^{5},
\end{align}    
\end{subequations}
where the centre of the plateau $x_{0}$ = 1 cm, the length of the plateau $l$ = 6 mm and the parameter $C_{r}$ = 80. Burke's reaction mechanism \cite{burke2012comprehensive} is employed as the chemical kinetics.

\begin{figure}[H]
\centering
    \centering
    \begin{subfigure}{1.0\textwidth}
        \centering
        \includegraphics[width=0.65\textwidth]{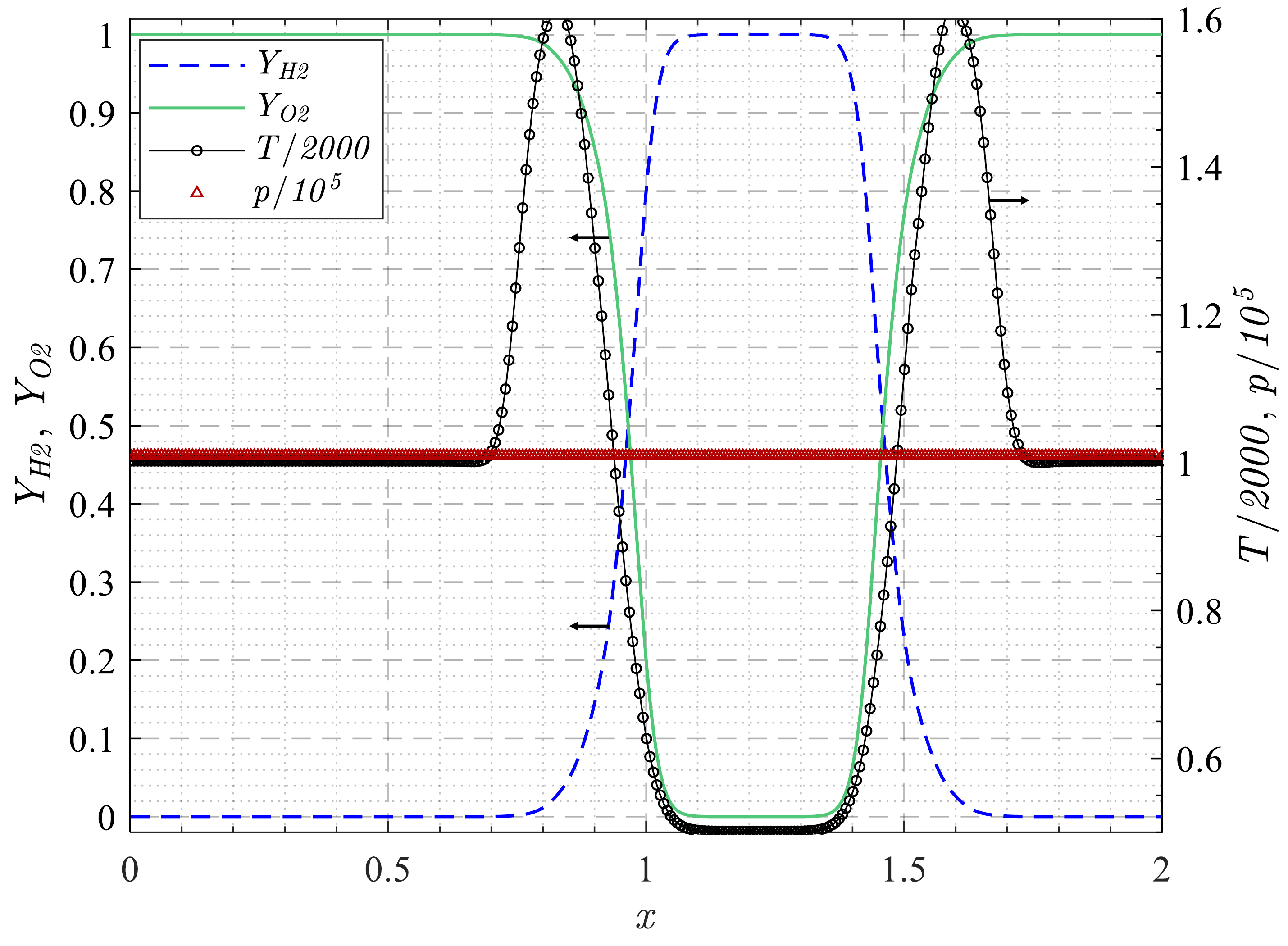}
        \caption{}
        \label{fig4-3-1(a)}    
    \end{subfigure}
    \begin{subfigure}{1.0\textwidth}
        \centering
        \includegraphics[width=0.65\textwidth]{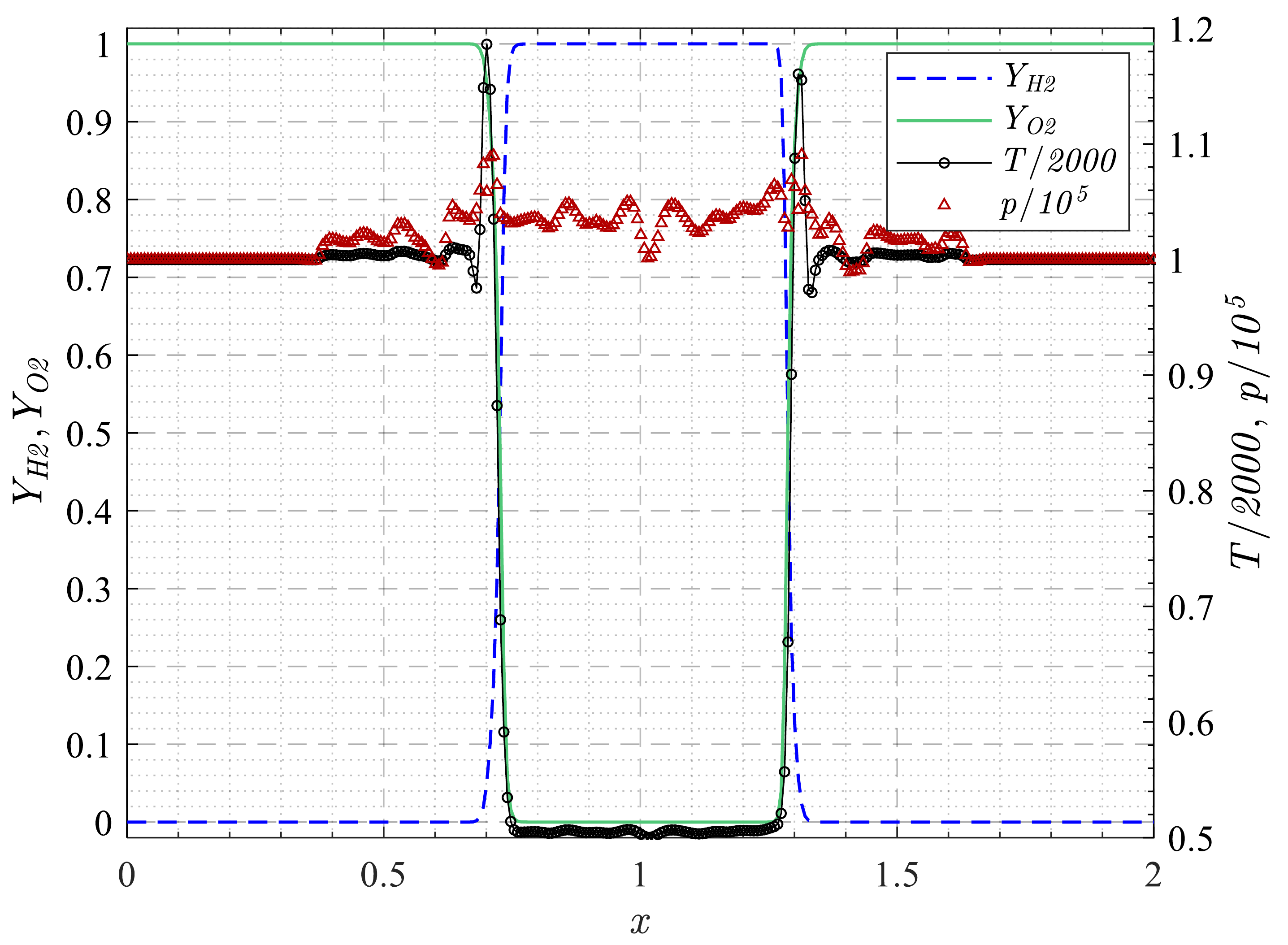}
        \caption{}
        \label{fig4-3-1(b)}    
    \end{subfigure}   
\caption{1-D reactive convection of hydrogen bubble solved with (a) extended double-flux scheme at $t_{1}$ = 0.1 ms and (b) fully conservative scheme at $t_{2}$ = 0.008 ms.}
\label{fig4-3-1}
\end{figure}

As shown in Fig. \ref{fig4-3-1(a)}, the extended double-flux scheme produces smooth profiles of the temperature $T$ as well as the mass fractions $Y_{\ce{H2}}$ and $Y_{\ce{O2}}$ until to the time $t_{1}$.
Although affected by the reaction process, the pressure exhibits negligible gradient at the reaction layers, possibly due to the dissipation of physical diffusion.
By contrast, as in Fig. \ref{fig4-3-1(b)}, the pressure at the bubble fronts oscillates significantly upon initiation of the calculation when using a fully conservative scheme, which results in the asymmetry of the temperature spikes and the oscillations nearby.
This kind of temperature oscillations at the material fronts might lead to a flawed description of the reaction process.
According to our results, with the same calculation time, the averaged values of the temperature spikes obtained with the extended double-flux scheme and the fully conservative scheme at $t=t_{1}$ are respectively 330.6 K and 347 K, which is 4.96$\%$ variation. 

\subsection{2-D DDT (Deflagration-to-Detonation Transition) in a smooth channel}
\label{sec4.7}
This example illustrates the capability of the new hybrid solver in handling premixed combustion phenomena efficiently and accurately, in the presence of strong shocks.
The calculation domain is a rectangle of length 20 cm by 2 cm, with 400$\times$40 cells in the $x$- and $y$-direction.
The case is initiated with quiescent, homogeneous $\ce{H2}$-$\ce{O2}$ premixed gas in the stoichiometric equivalent ratio, characterized with a Lewis number of $Le_{\text{eff}}$ = 0.8.
Table \ref{tab3} lists the input parameters as well as the resulting flame properties obtained from Cantera \cite{goodwin2018cantera}.
\begin{table*}[h]
\centering
\footnotesize
\caption{Input parameters and output thermal properties for hydrogen-oxygen mixture at the equivalent ratio. $\delta_{\text{T}}=(T_{\text{ad}} - T_0)/|\text{d}T/\text{d}x|_{\text{max}}$.}
\renewcommand\arraystretch{1.2}
\begin{tabular}{lll}
\hline

\multicolumn{3}{l}{Inputs Parameters}   
\footnotesize
\\ \hline
$p_0$        & 0.1 MPa                       & Initial pressure                      \\
$T_0$        & 300 K                         & Initial temperature                   \\
$\gamma$     & 1.4                           & Specific heat ratio                   \\
$\mu_0$     & 7.61e-05 kg/(m$\cdot$s)             & Mixture-averaged viscosity            \\
$\lambda_0$ & 1.875e-03 m$^{2}$/s                & Thermal diffusivity \\ 
\addlinespace[5pt] 
\hline
\multicolumn{3}{l}{Flame Properties}                                                              \\ \hline
$Le_{\text{eff}}$   & 0.8                         & Effective Lewis number                \\
$\delta_T$  & 2.59e-04 m                    & Laminar flame thickness               \\
$S_L$      & 9.41 m/s                      & Laminar flame speed                   \\
$T_{\text{ad}}$    & 3071 K                    & Adiabatic flame temperature        \\
$\Theta$     & 10.24                          & Thermal expansion ratio                       \\ \hline
\end{tabular}
\label{tab3}
\end{table*}
The mixture is ignited initially at the end wall with an ignition radius of $r$ = 0.1 cm, and then the flame propagates from the closed end to the open end in an increasing speed due to the well-known thermal-shock interaction \cite{lee2008detonation}.
All boundaries except for the right one are set as no-slip solid walls while the right boundary is treated as an extrapolation outlet.

\begin{figure}[H]
    \centering
    \begin{subfigure}{0.48\textwidth}
        \centering
        \includegraphics[width=\textwidth]{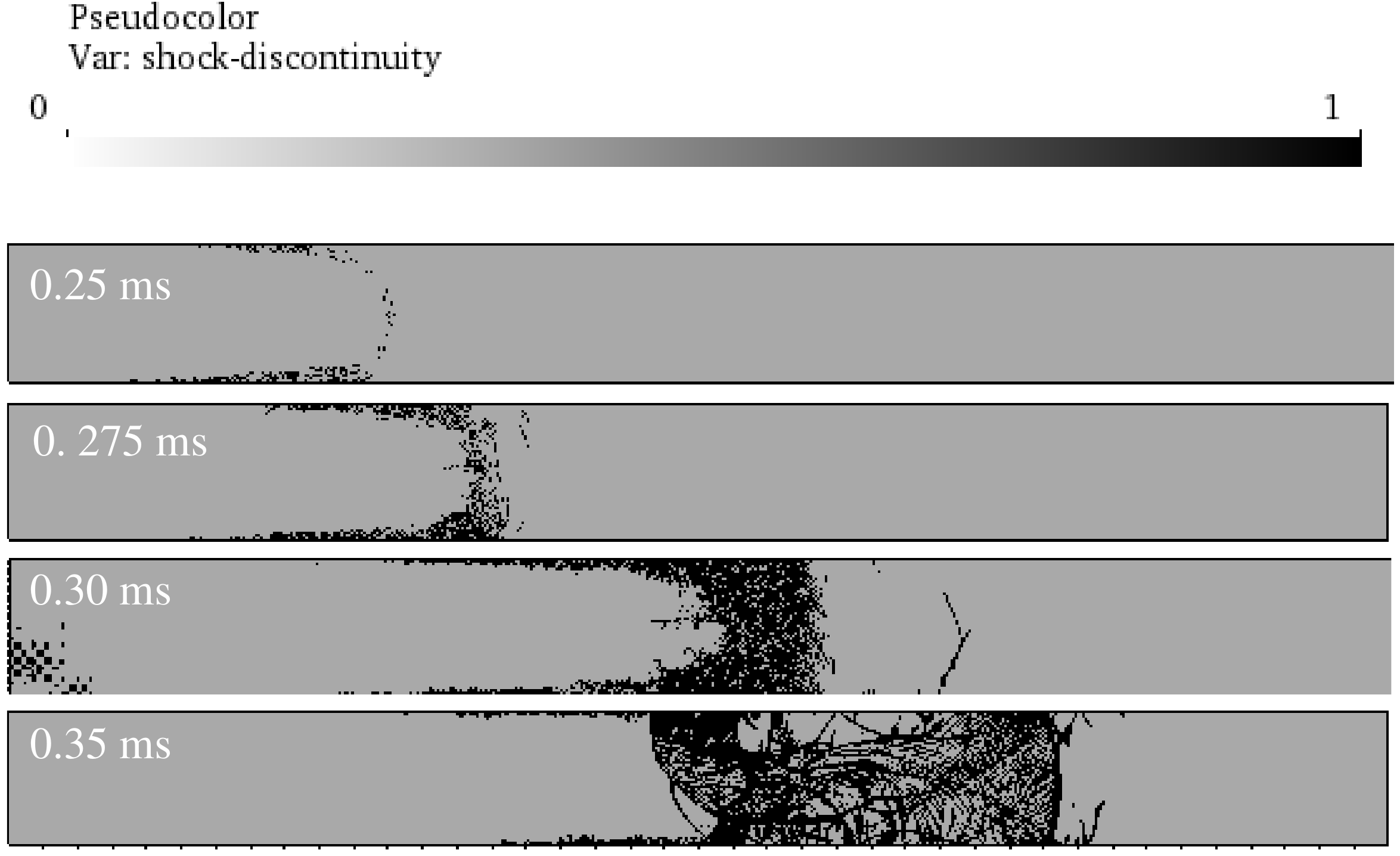}
        \caption{shock-discontinuity flagging}
        \label{fig4-7-1a}
    \end{subfigure}
    \hfill 
    \begin{subfigure}{0.48\textwidth}
        \centering
        \includegraphics[width=\textwidth]{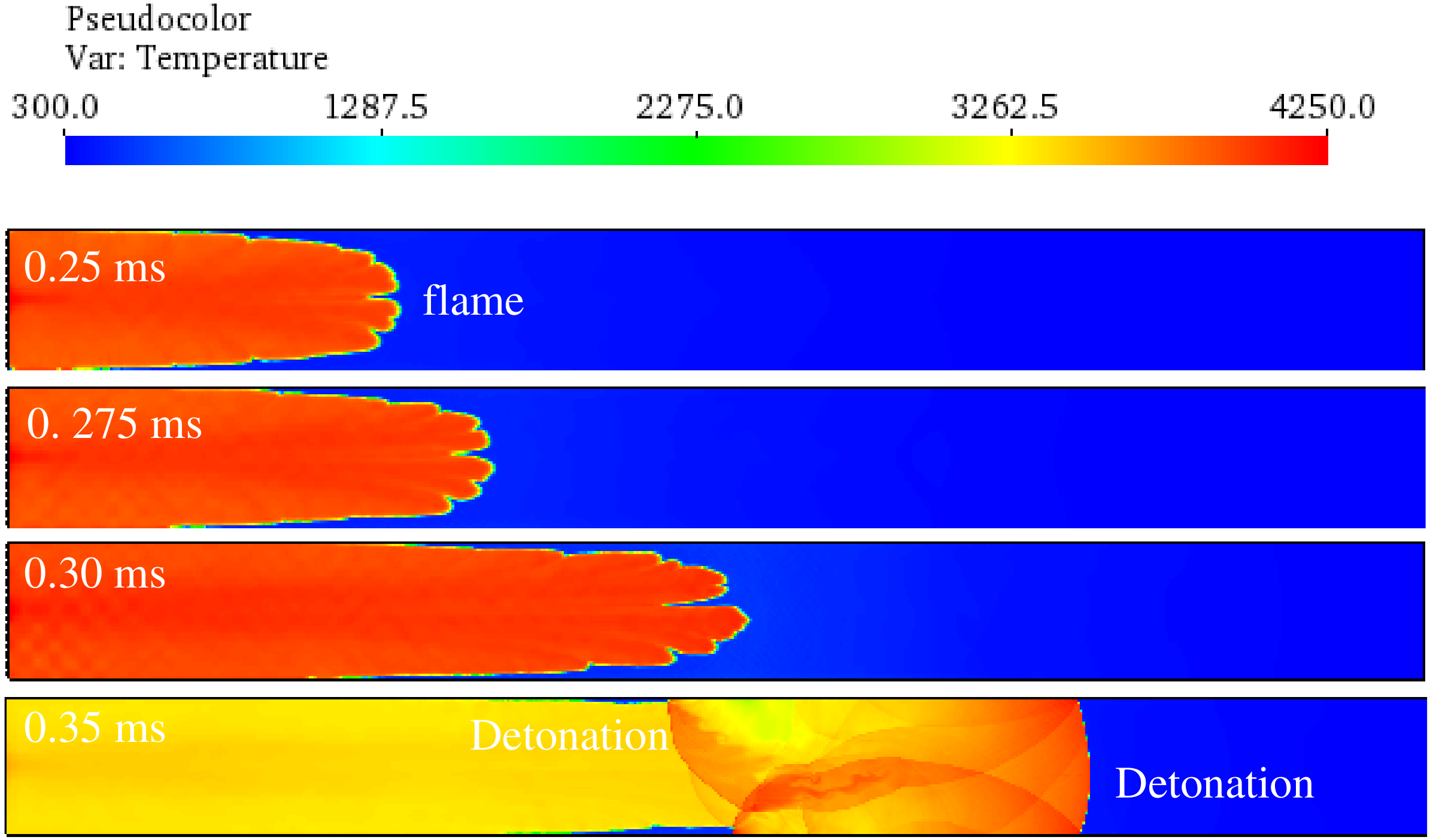}
        \caption{temperature}
        \label{fig4-7-1b}
    \end{subfigure}
    \vspace{1em} 
    \begin{subfigure}{0.48\textwidth}
        \centering
        \includegraphics[width=\textwidth]{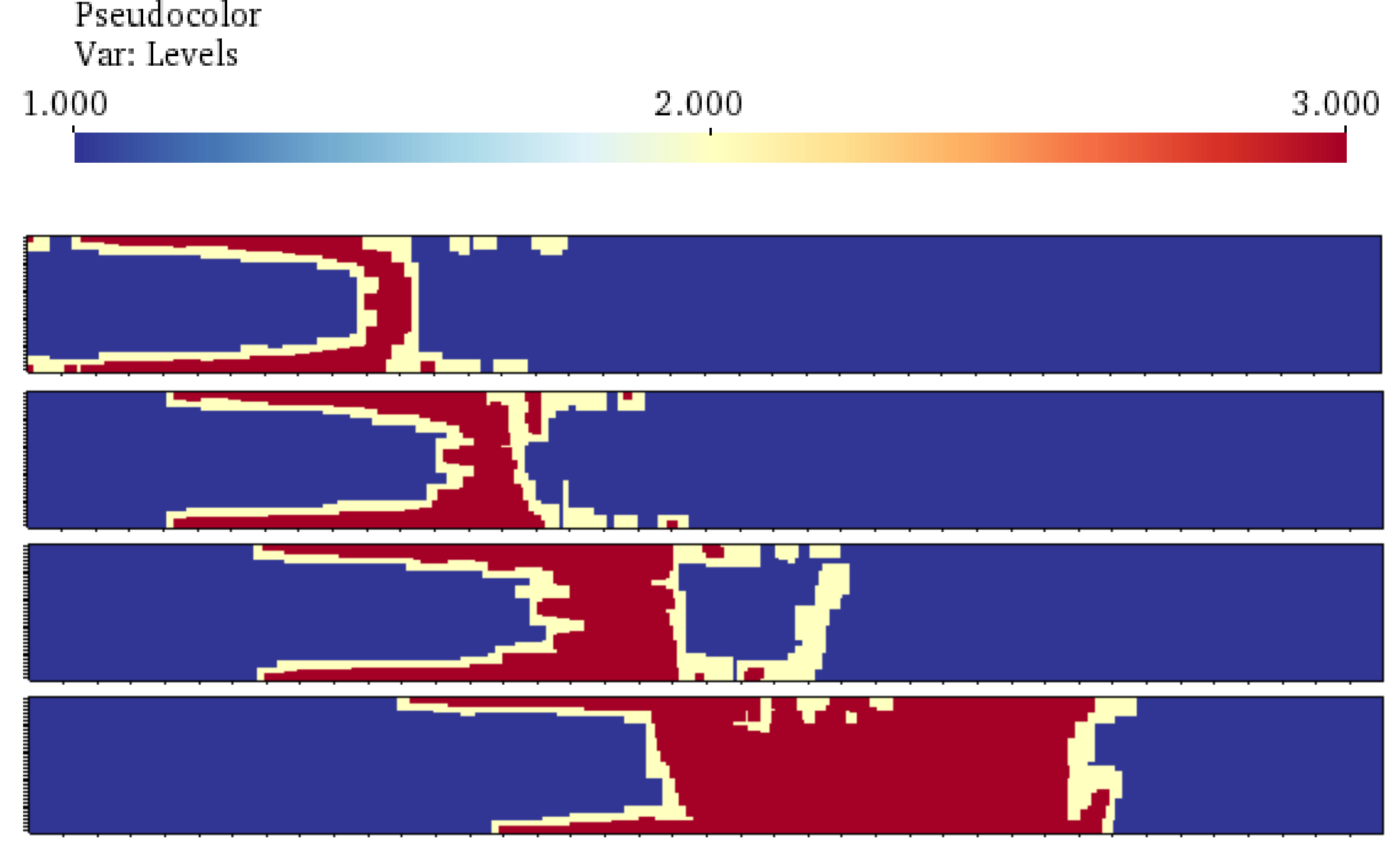}
        \caption{refinement level}
        \label{fig4-7-1c}
    \end{subfigure}
    \caption{Illustration of numerical flame acceleration and detonation initiation process with 3-level refinement (2,4) in case 1 by (a) shock-discontinuity flagging, (b) temperature and (c) refinement level diagrams.}
    \label{fig4-7-1}
\end{figure}

We simulate two cases with 3 and 4 refinement levels respectively, as listed in Table \ref{tab4}. 
\begin{table*}[h]
\centering
\footnotesize
\caption{Refinement configuration of case 1 and case 2 in the premixed DDT example.}
\begin{tabular}{ccc}
\hline
                                 & case 1        & case 2         \\ \hline
refinement levels $l$            & 3            & 4             \\
refinement ratio $r$             & (2,4)        & (2,2,4)       \\
minimum grid size $\delta_{min}$ & 62.5 $\mu m$ & 31.25 $\mu m$ \\ \hline
\end{tabular}
\label{tab4}
\end{table*}
Results obtained with the 3-level refinement are shown  in Fig. \ref{fig4-7-1}.
The snapshots of the shock flagging criteria as well as the temperature diagrams during the dynamic DDT process illustrate that the hybridization described in Section \ref{sec3.2} worked quite well, identifying most of the important discontinuous phenomena such as the compression waves, the precursor shock as well as the detonation wave.
To further assess the efficiency of the presented new hybrid solver in handling higher grid-level refinement, which is usually required for high resolution DDT simulation, we continue investigating the same configuration with a finer grid of a 4-level refinement.  

\begin{figure}[H]
    \centering
    \includegraphics[width=0.5\textwidth]{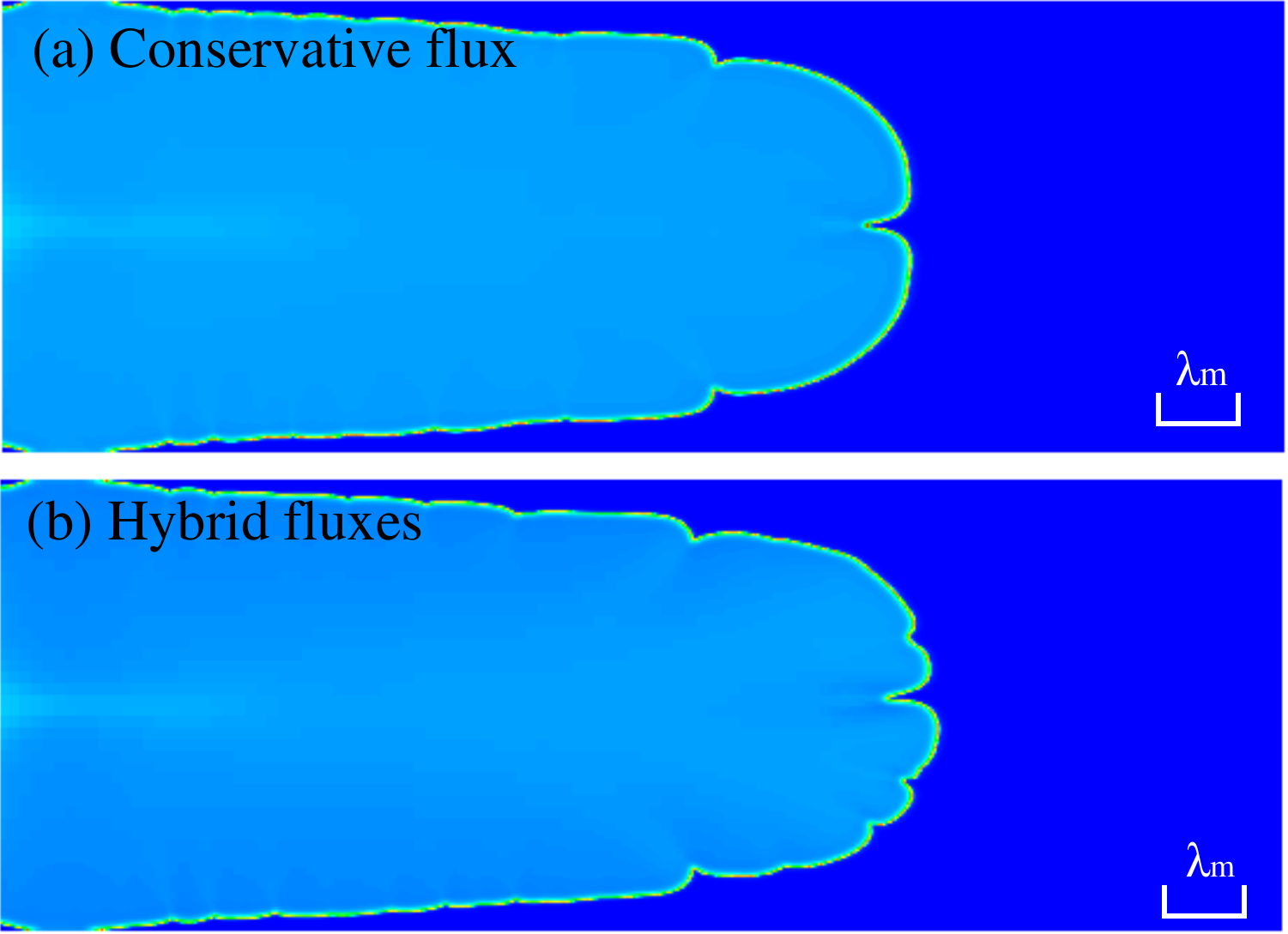}
    \caption{Unstable cellular structures solved respectively by the fully conservative scheme and the hybrid scheme with 4-level refinement.}
    \label{fig4-7-2}
\end{figure}

Figure \ref{fig4-7-2} illustrates the fine structures of an unstable flame front, obtained from the fully conservative and hybrid schemes with 4-level refinement respectively.
As can be observed in Fig. \ref{fig4-7-2}(a), the flame front solved with the fully conservative scheme evolves into a single-cusp structure with a cell size approximately equal to $h$.
In contrast, a multi-head cellular structure is established at the flame front in Fig. \ref{fig4-7-2}(b) solved by the hybrid solver, which aligns more with the most unstable wavelength $\lambda_{m}$ = 6$\delta_{T}$ as predicted by linear flame instability analysis \cite{altantzis2012hydrodynamic,berger2019characteristic}.
The length of $\lambda_{m}$ is also labelled in Fig. \ref{fig4-7-2} for comparison.
We analyze this difference to be attributed to the transverse propagation of the spurious oscillation waves in Fig. \ref{fig4-7-2}(a), which causes the numerical merging of adjacent small cusps.
For the hybrid counterparts, on the other hand, the material interfaces remain free of oscillation and thus reproduce the real-world flame evolution better.

\begin{figure}[H]
    \centering
    \includegraphics[width=0.6\textwidth]{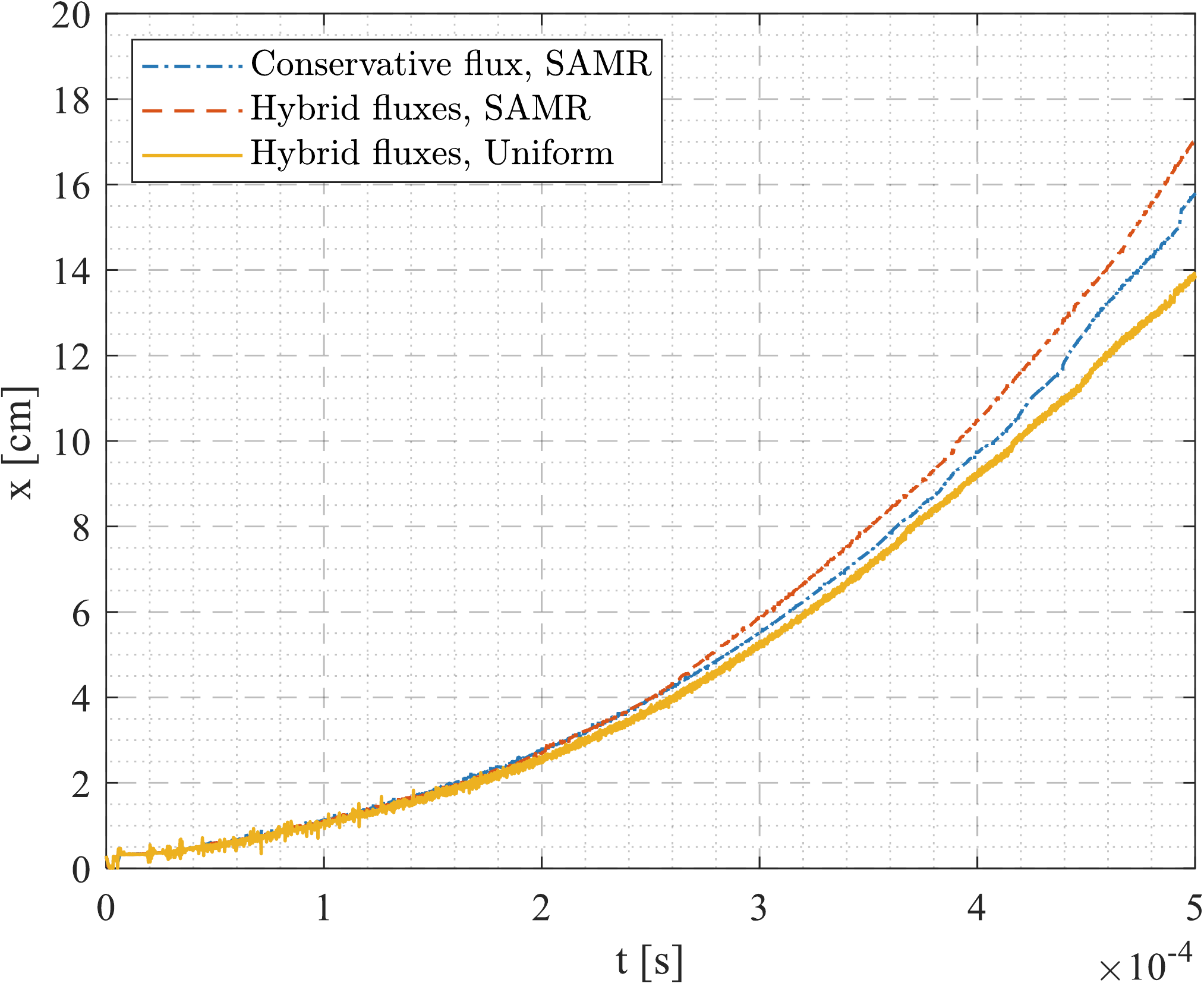}
    \caption{Comparison of time-space flame front trajectories obtained using the fully conservative scheme and extended double-flux scheme with SAMR, as well as the one obtained using a uniform fine grid.}
    \label{fig4-7-3}
\end{figure}

\begin{figure}[H]
    \centering
    \includegraphics[width=0.6\textwidth]{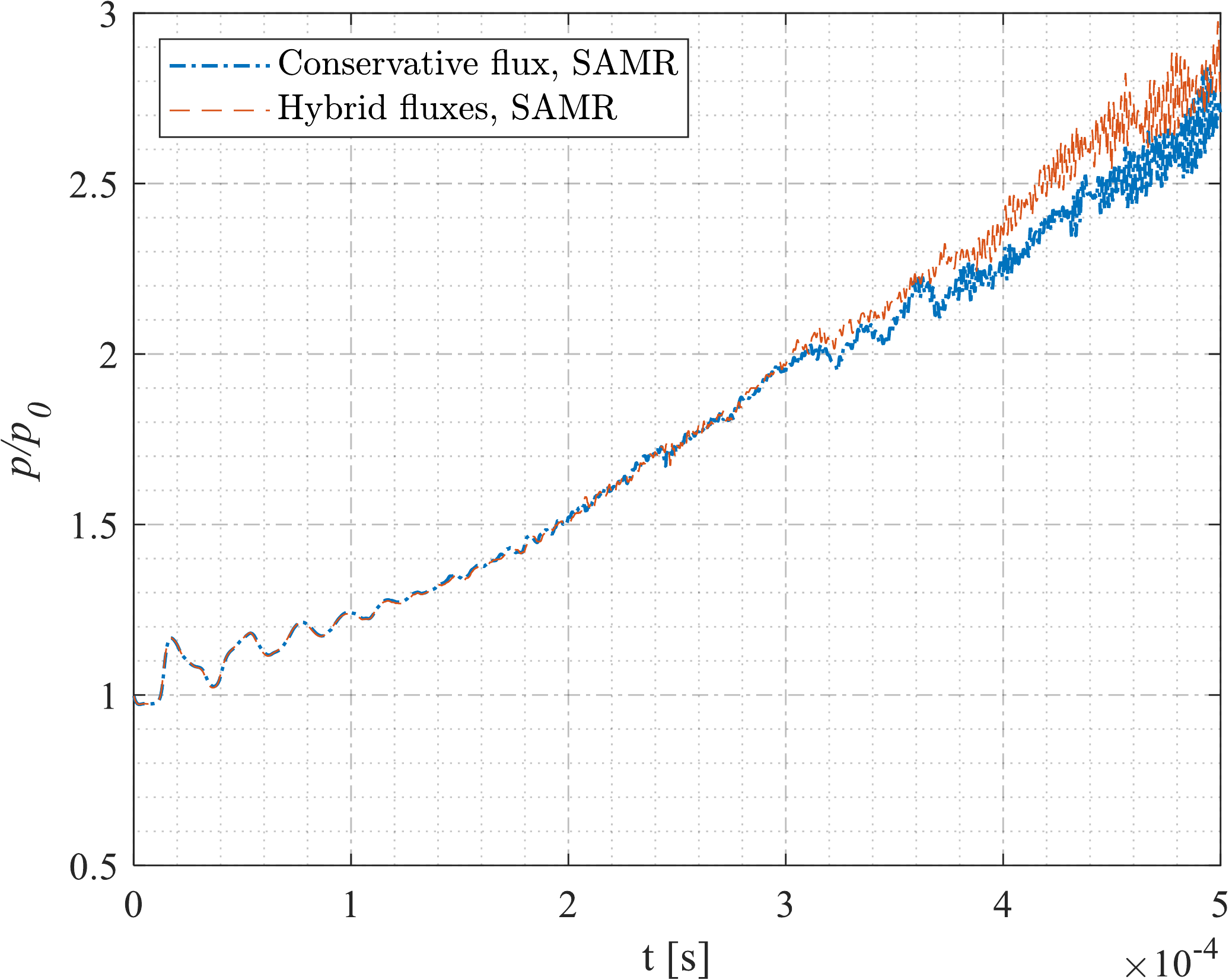}
    \caption{Comparison of the end-wall pressure history obtained using the fully conservative scheme and the hybrid scheme with SAMR.}
    \label{fig4-7-4}
\end{figure}

In general, Figs \ref{fig4-7-3} and \ref{fig4-7-4} both illustrate that the usage of the double-flux method generates a more accelerated flame as compared to the one solved with the conventional scheme.
That is reasonable because a sharper folding flame profile obtained means more heat release onto the leading shock that will compress the reactants behind the flame.
This is also evidenced by the early agreement of the two curves in Figs.  \ref{fig4-7-3} and \ref{fig4-7-4} during which the separate two flames are more dominated by initial thermal expansion instead of flame instabilities.
In Fig. \ref{fig4-7-3}, we also compare the result solved by a uniform grid with the grid size $\delta _{u}$ = 31.25 $\mu m$, which corresponds right to the finest grid size $\delta_{min}$ in the 4-level SAMR configuration.
Generally, the result is in agreement with the one solved by the SAMR, illustrating the SAMR working well.
Table \ref{tab5} compares the total wall cpu time for a run on 400 cores, which proves again the efficiency of our presented hybrid method as well as the methodology by combining the hybrid solver with the SAMR technique.

\begin{table}[H]
\centering
\footnotesize
\caption{Comparison of the wall cpu time calculated by a 4-level SAMR grid and a uniform grid with the grid size corresponding to the finest one in the 4-level SAMR grid.}
\label{tab5}
\begin{tabular}{cccc}
\hline
                      & \multicolumn{2}{c}{SAMR Grids,}                & \multirow{2}{*}{Uniform Grids} \\
                      & \multicolumn{2}{c}{4-level refinement (2,2,4)} &                                \\ \cmidrule(lr){2-3} \cmidrule(lr){4-4} 
                      & conservative scheme       & hybrid scheme      & hybrid scheme                  \\ \hline
wall cpu time {[}s{]} & 10429.475                 & \color{red}10920.522          & 27393.394                      \\ \hline
\end{tabular}
\end{table}

\subsection{2-D non-premixed planar RDE (Rotating Detonation Engine)}
\label{sec4.8}
In this section, a detonation wave flowing over $\ce{H2}$/air slots on a planar wall, i.e., a 2D non-premixed planar RDE configuration, is simulated.
The computational domain is a rectangle with $n_{x}\times n_{y}$ = 100 $\times$ 400 cells in the $x$- and $y$-direction.
The domain lengths in the $x$- and $y$-directions are $l_{x}$=5 cm and $l_{y}$=20 cm, respectively.

As for the boundary conditions, both of the upper and lower ends of this computational domain are treated as periodic.
At the left end, a total of $N$=40 injection nozzles are distributed uniformly along the length at an interval of $d_{1}$=3 mm with non-premixed $\ce{H2}$ and air injected separately in their sonic speeds.
The width of each slot is $d_{2}$=2 mm.
In this non-premixed scenario, the injection width of $\ce{H2}$ is 1/4 of the whole slot width and so the air fills the remaining 3/4 part in each slot.
During each interval, a slip wall condition is used.
In this example, the injection time depends on the relation of the total pressure of the non-premixed gases $p_{\text{tol}}$ and the dynamic interior pressure $p_{\text{in}}$ due to wave propagation.
When $p_{\text{tol}} \ge p_{\text{in}}$, the injection is initiated, otherwise the slots are treated numerically as slip walls.
The injection total pressure and temperature are set to $p_{\text{tol}}$ = 0.5 MPa and $T_{\text{tol}}$ = 300 K, respectively.
At the right end of the domain, a hybrid subsonic/supersonic outflow boundary is applied.

\begin{figure}[H]
    \centering
    \includegraphics[width=0.6\textwidth]{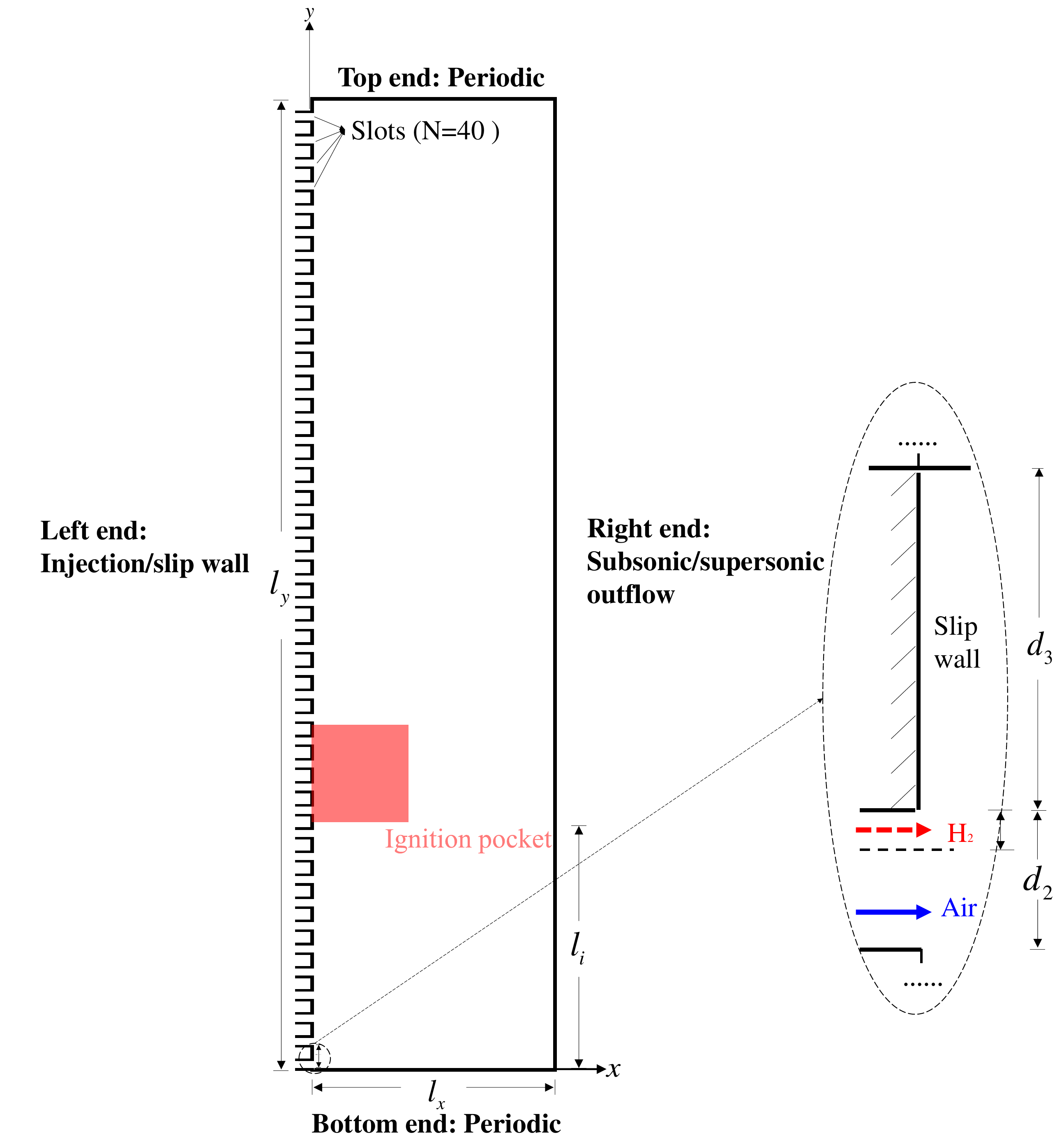} 
    \caption{Numerical configuration of the 2-D planar RDE example.}
    \label{fig4-8-1}
\end{figure}

\begin{figure}[H]
    \centering
    \includegraphics[width=0.9\textwidth]{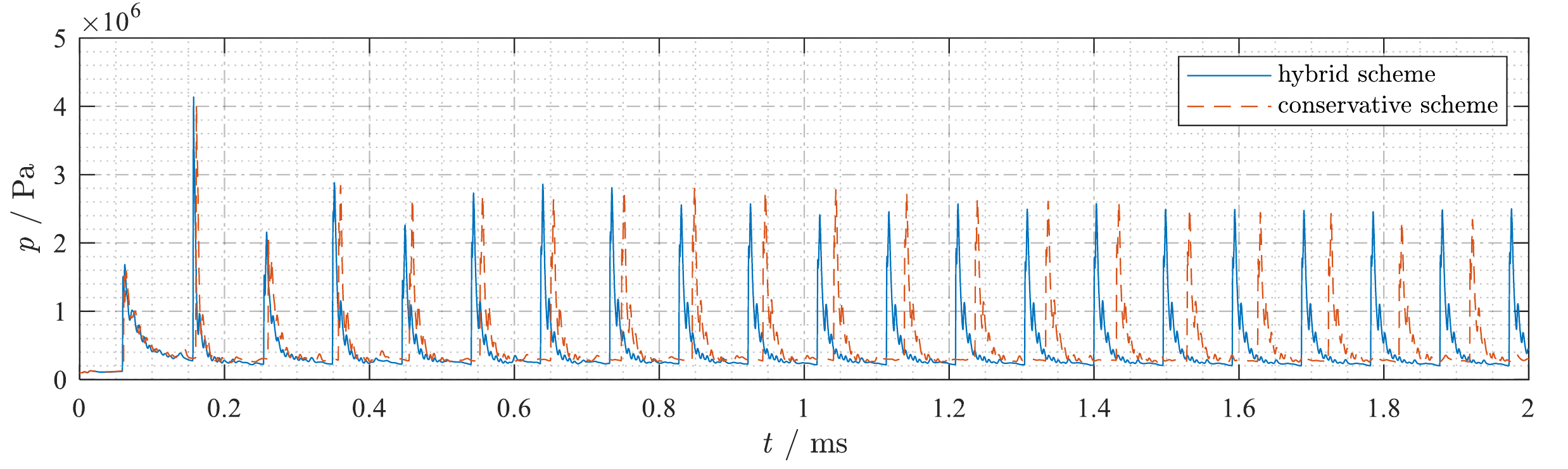} 
    \caption{Pressure time history at line x = 1 cm solved by the conservative and hybrid schemes.}
    \label{fig4-8-2}
\end{figure}

This example is initiated by placing an ignition hot pocket onto the reactive mixtures, as depicted in Fig. \ref{fig4-8-1}.
Owing to the use of periodic boundary condition, the detonation propagates periodically within the computational domain.
The total simulation time is 2 ms, during which the detonation cycle occurs about 21 times.
It can be observed from Fig. \ref{fig4-8-2} that except from the first detonation cycle, divergence of the detonation propagation trajectories solved by the fully conservative and hybrid schemes is continuously increasing.
The first detonation cycle, on the other hand, remains unaffected mainly due to the imposition of the initial condition during which the injection and mixing processes have had no effects yet.

\begin{figure}[H]
    \centering
    \begin{subfigure}{0.48\textwidth}
        \centering
        \includegraphics[width=\textwidth]{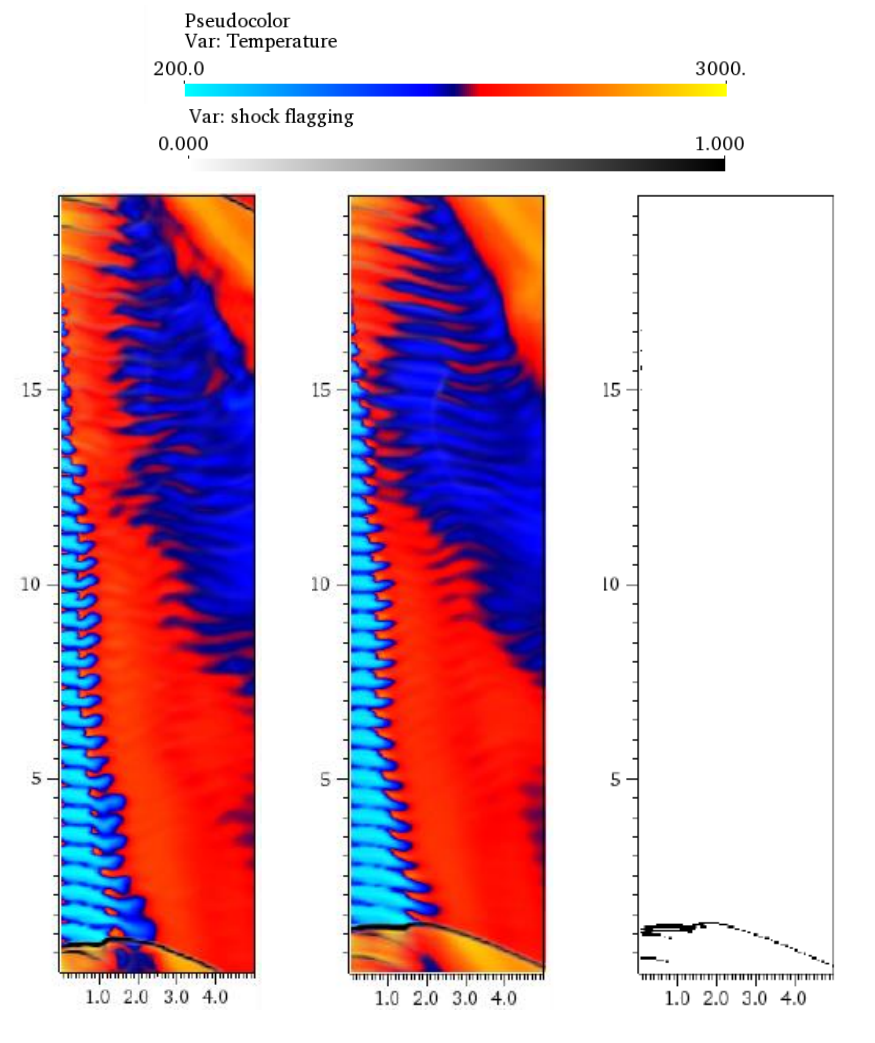}
        \caption{$t$ = 0.1625 ms}
        \label{fig4-8-3a}
    \end{subfigure}
    \hfill 
    \begin{subfigure}{0.48\textwidth}
        \centering
        \includegraphics[width=\textwidth]{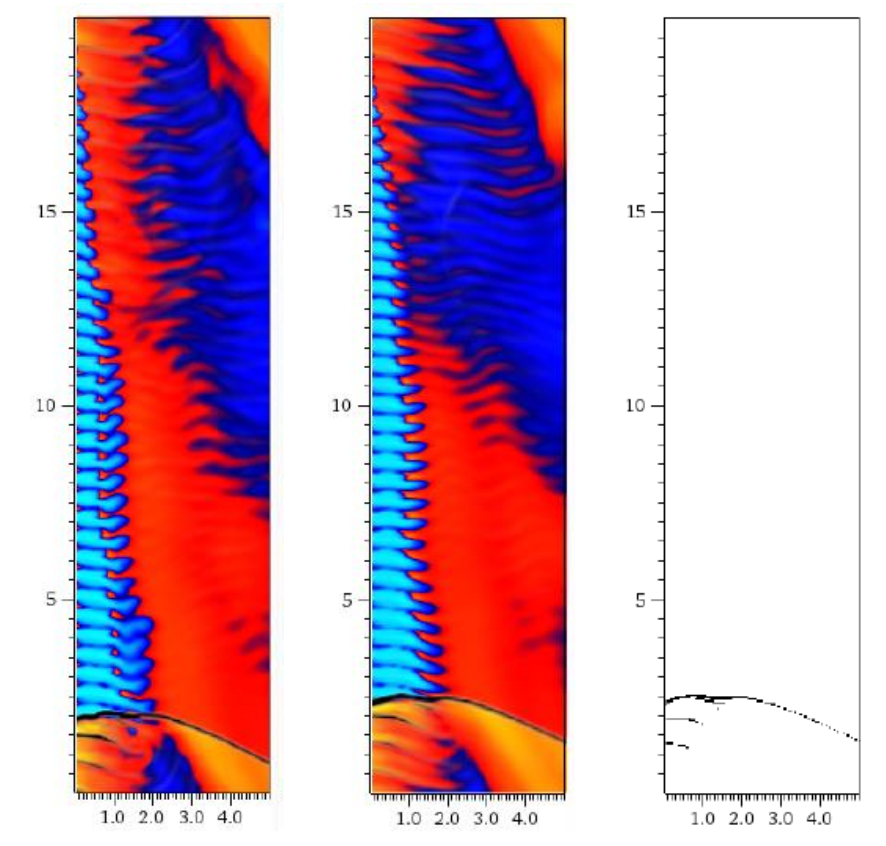}
        \caption{$t$ = 0.1692 ms}
        \label{fig4-8-3b}
    \end{subfigure}
    \vspace{1em} 
    \begin{subfigure}{0.48\textwidth}
        \centering
        \includegraphics[width=\textwidth]{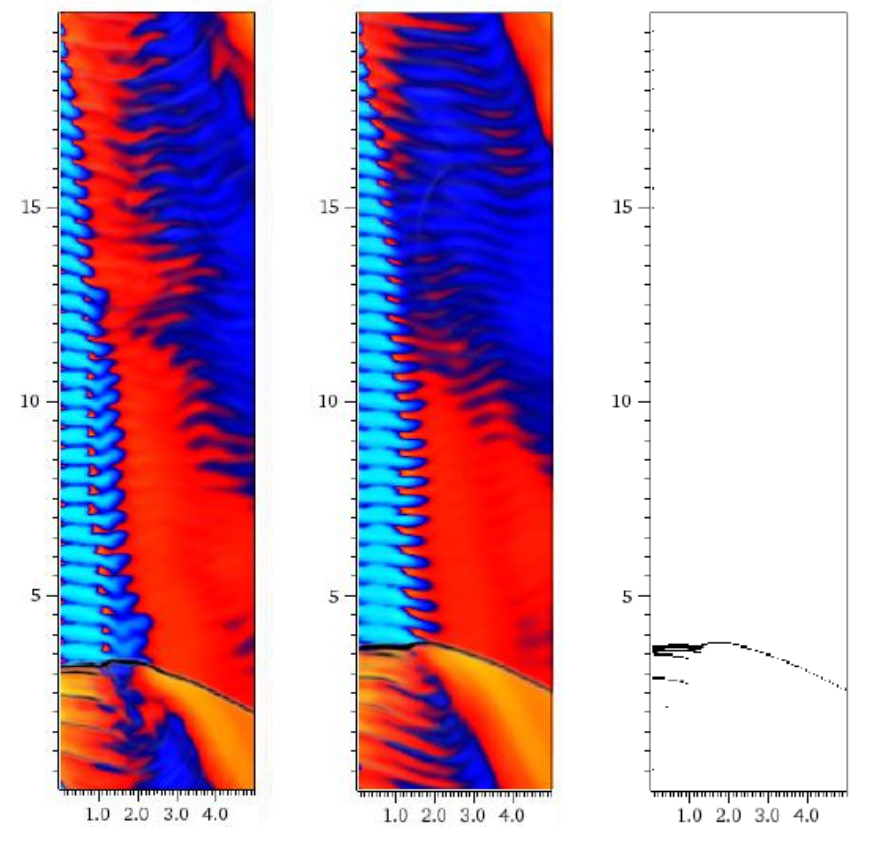}
        \caption{$t$ = 0.1758 ms}
        \label{fig4-8-3c}
    \end{subfigure}
    \hfill 
    \begin{subfigure}{0.48\textwidth}
        \centering
        \includegraphics[width=\textwidth]{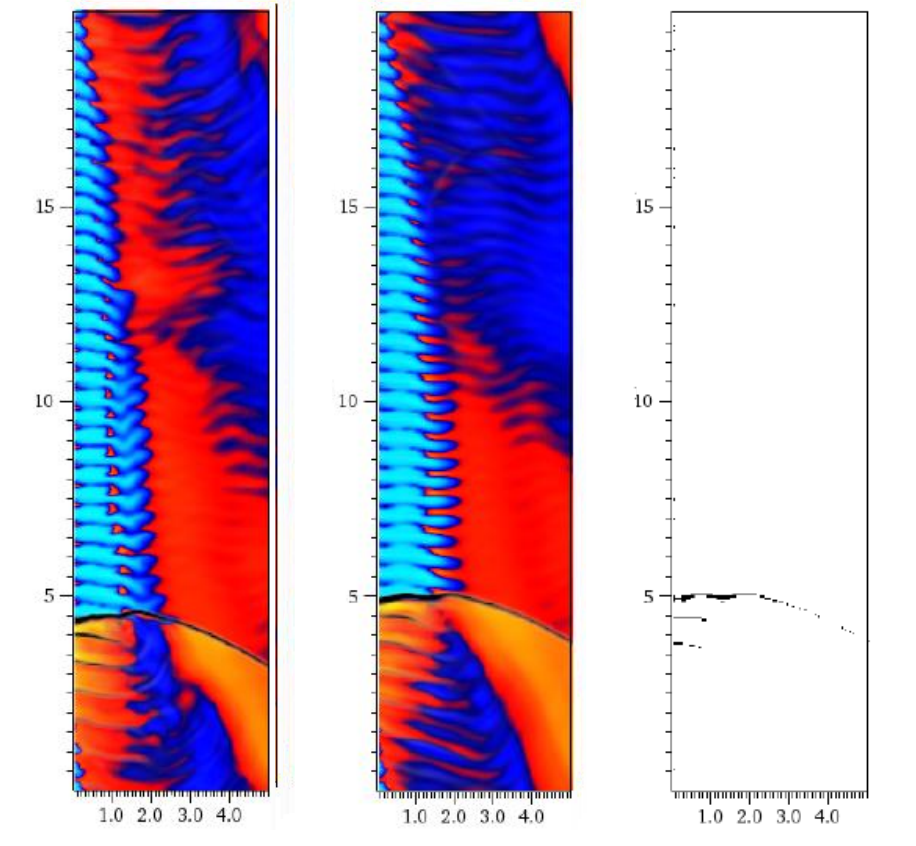}
        \caption{$t$ = 0.1825 ms}
        \label{fig4-8-3d}
    \end{subfigure}
    \caption{Comparison of dynamic injection and detonation wave propagation calculated by the conservative and hybrid schemes. Left column: conservative scheme; Middle column: hybrid scheme; Right column: shock flagging.}
    \label{fig4-8-3}
\end{figure}

To illustrate this divergence, the beginning stage of the second detonation cycle is shown in Fig. \ref{fig4-8-3}, which depicts a typical temperature and shock flagging fields during this stage.
In the simulation result from the conservative scheme, a partial extinction phenomenon can be seen clearly in Figs. \ref{fig4-8-3a} to \ref{fig4-8-3d} that occurs far from the injection slots and where the detonation front merges with the previous burned products . As the detonation wave travels further, the extinction becomes more severe. On the other hand, the simulation result from the new hybrid scheme shows clearly that the detonation wave as well as the detached compression waves behind it are well-flagged with the shock sensor presented in Section \ref{sec3.2}.
Meanwhile, in contrast with the result obtained from the conservative scheme, it can also be observed that the head of the reaction wave is continuously coupled with the precursor shock, or else, the detonation is self-sustained during its propagation.  

\begin{figure}[H]
\centering
    \begin{subfigure}{0.5\textwidth}
        \includegraphics[width=\textwidth]{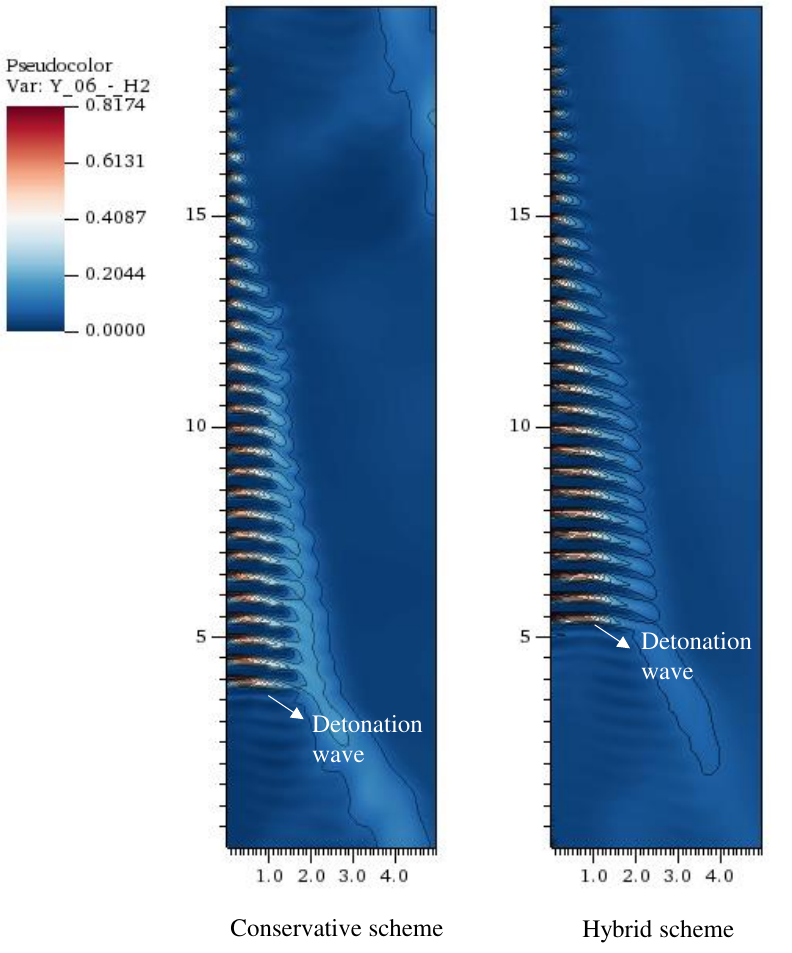}
        \caption{mass fraction of \ce{H2} fields}
        \label{fig4-8-5a}    
    \end{subfigure}
    \hfill %
    \begin{subfigure}{0.4\textwidth}
        \includegraphics[width=\textwidth]{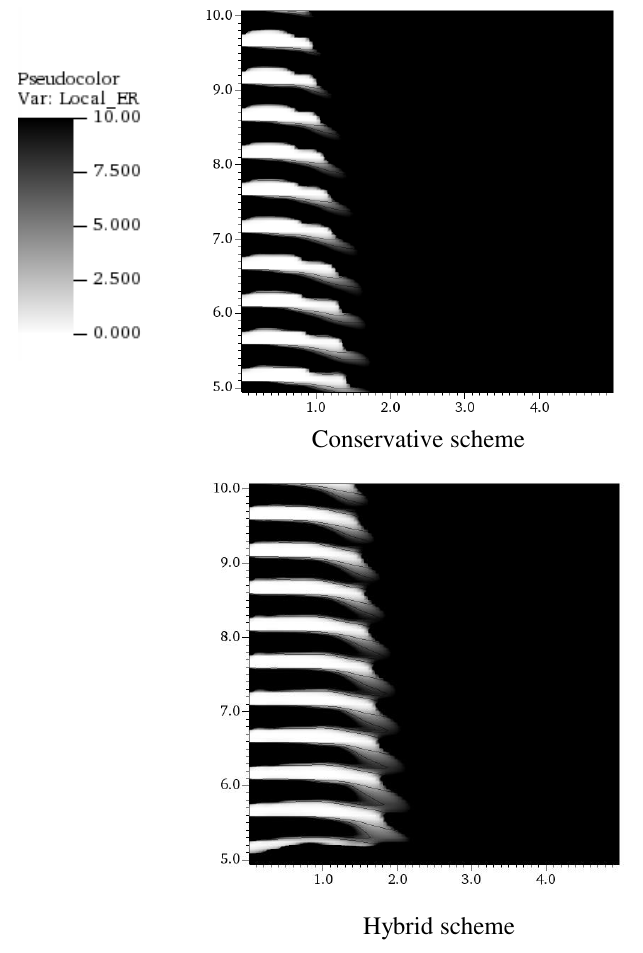}
        \caption{local ER fields}
        \label{fig4-8-5b}    
    \end{subfigure}
\caption{Comparison of the (a) mass fraction of \ce{H2} global fields, (b) local equivalence ratio (ER) local fields calculated by the conservative and hybrid schemes.}
\end{figure}

To uncover the underlying mechanism of such a significant divergence, we analyze the mixing fields derived respectively from the conservative and presented hybrid schemes, represented by the mass fraction of $\ce{H2}$ in Fig. \ref{fig4-8-5a}.
It can be observed that the mixing region solved with the conservative scheme is much more oscillatory than that solved with the presented hybrid scheme.
This kind of spurious oscillation interrupts the regular injection mixing process, and thus pollutes the mixing regions.
By taking the $x$-axis direction as the reference direction, this interaction between the transverse oscillation waves and the longitudinal injection mixing would cause such a huge mixing flaw, especially in such a device with multiple injection slots and multiple injection mixing occurring simultaneously.

To take a deeper look at this phenomena from Fig. \ref{fig4-8-5b}, we could finally conclude that the partial detonation extinction essentially originates from its local high equivalence ratio (ER).
The presented hybrid solver performs better to resolve the injection mixing process and thus derives a more suitable low-ER mixing region for detonation sustainability.
Therefore, we can conclude that when the traditional conservative scheme is utilized on a non-premixed combustion scenario, the resulting impact is can be quite catastrophic, which fortunately can be cured by using the hybrid scheme presented in this paper.

\section{Conclusion}
\label{sec5}
The present study has successfully advanced a hybrid methodology within the finite-volume method, achieving a verified precision of second-order accuracy. The proposed new hybrid solver merges an extended double-flux strategy with conventional conservative methodologies, thereby ensuring enhanced stability and accuracy in multi-dimensional scenarios.

Central to this solver are the carefully tailored hybrid schemes for managing convective fluxes and interpolations. These designs guarantee optimal performance in maintaining solution stability while simultaneously delivering high-fidelity results, even under challenging conditions that combine material interfaces with shock discontinuities.

To extend the original double-flux technique, this paper put forth two full computational approaches, which have both proven reliable and efficient in suppressing pressure oscillations occurring at the material interfaces and at the same time require only one auxillary variable $\hat{\gamma}$, with tiny difference in momentum conservation preservation.

Extensive validations have conclusively demonstrated the efficiency and resilience of our solver in tackling multi-dimensional, shock-interface challenges, even in scenarios complicated by viscosity and chemical reactions by combining with the adaptive mesh refinement (AMR) technique. 
In the future, we expect to leverage the solver to handle large-scale compressible chemically reacting problems.

\clearpage %
\appendix

\section{Hybridized solution procedure}
\label{ap0}

\begin{algorithm}[!htb]
\caption{Hybridization methodology}
\label{al1}
\begin{algorithmic}[1]
\STATE Calculate and store cell-based auxiliary variable $\hat{\gamma}$ using either Eq. \ref{eq27} or Eq. \ref{eq32}.
\STATE Compute the troubled-cell indicator $S_{k}$ for shock detection ($> S_{k}^{T}$ using shock-capturing scheme, $> S_{k}^{T}$ with extended double-flux scheme).
\IF {$S_{k} < S_{k}^{T}$}
  \STATE Save $\boldsymbol{U^n}$ first
  \STATE \textit{First RK Stage:}
  \STATE Interpolate primitive variables onto cell edges; calculate inconsistent numerical fluxes using the extended double-flux scheme with the addition of viscous fluxes
  \STATE Update state vector $\boldsymbol{U^n}$ with dual edge fluxes to obtain $\boldsymbol{U^*}$.
  \STATE Update primitive variables as follows:
  \STATE \hspace{1em} $Y_i^* = \frac{(\rho Y_i)^*}{\sum (\rho Y_i)^*}$, $u^* = \frac{(\rho u)^*}{\sum (\rho Y_i)^*}$, $v^* = \frac{(\rho v)^*}{\sum (\rho Y_i)^*}$
  \STATE \hspace{1em} $p^* = \left(\hat{\gamma}-1\right)\left((\rho E)^* - (\rho h_{0}^f)^* - \frac{\rho^* \boldsymbol{u}^{*2}}{2}\right)$, $T^* = \frac{p^*}{RU\sum \frac{(\rho Y_i)^*}{W_i}}$
  \STATE \textit{Second RK Stage:}
  \STATE Repeat steps 6-10 for updating of state vector $\boldsymbol{U^*}$ to get the final state $\boldsymbol{U^{n+1}}$.
  \STATE Update thermodynamic variables:
  \STATE \hspace{1em} $(C_{pi}^*)^{n+1} = C_{pi}^*(T^{n+1})$, $(\hat{\gamma})^{n+1} = \frac{C_{p}^*}{C_{p}^*-R}$
  \STATE Correct the total energy $(\rho E)^{n+1}$:
  \STATE \hspace{1em} $(\rho E)^{n+1} = \frac{p^{n+1}}{(\hat{\gamma})^{n+1} -1} + (\rho h_0^f)^{n+1} + \frac{(\rho \boldsymbol{u}^2)^{n+1} }{2}$
\ELSE
  \STATE Save $\boldsymbol{U^n}$ first
  \STATE \textit{First RK Stage:}
  \STATE Interpolate charatertistic variables onto cell edges; calculate conservative numerical fluxes using the conservative scheme with the addition of viscous fluxes
  \STATE Update state vector $\boldsymbol{U^n}$ with single edge flux to obtain $\boldsymbol{U^*}$.
  \STATE \textit{Second RK Stage:}
  \STATE Repeat steps 20-21 for updating of $\boldsymbol{U^*}$ to get the final state $\boldsymbol{U^{n+1}}$.
\ENDIF

\end{algorithmic}
\end{algorithm}

\section{Left and right characteristic matrix of multicomponent Navier-Stokes equations}
\label{apA}
The Jacobian matrix $
 \mathrm{A}_{1}(\mathrm{q})$ has a complete set of eigenvectors, hence it is diagonalizable with  $\mathbf{R}_{1}^{-1}(\mathbf{q}) \mathbf{A}_{1}(\mathbf{q}) \mathbf{R}_{1}(\mathbf{q}) = \boldsymbol{\Lambda}_{1}(\mathbf{q})$  for all admissible states $\mathbf{q}$  with $\boldsymbol{\Lambda}_{1}(\mathbf{q})= \operatorname{diag}\left(u_{1}-c, u_{1}, \ldots, u_{1}, u_{1}+c\right)$.
 In the matrix of right eigenvectors 
\scalebox{0.8}{}
\begin{align}
\mathbf{R}_{1}(\mathbf{q}) & = \left[\begin{array}{cccccccc}
Y_{1} & 1 & 0 & \ldots & 0 & 0 & 0 & Y_{1} \\
\vdots & 0 & & \ddots & \vdots & \vdots & \vdots & \vdots \\
Y_{K} & 0 & \ldots & 0 & 1 & 0 & 0 & Y_{K} \\
u_{1}-c & u_{1} & \ldots & u_{1} & 0 & 0 & u_{1}+c \\
u_{2} & u_{2} & \ldots & u_{2} & 1 & 0 & u_{2} \\
u_{3} & u_{3} & \ldots & u_{3} & 0 & 1 & u_{3} \\
H-u_{1} c & \mathbf{u}^{2}-\frac{\phi_{1}}{\bar{\gamma}} & \ldots & \mathbf{u}^{2}-\frac{\phi_{K}}{\bar{\gamma}} & u_{2} & u_{3} & H+u_{1} c
\end{array}\right]
\end{align}
the first and the last column are the vectors that span up the one-dimensional vector spaces corresponding to $\textit{u}_1-\textit{c}$ and $\textit{u}_1+\textit{c}$.
The vectors of the columns 2 to $\textit{K}$+3 span up the $\textit{K}$+2-dimensional vector space for $\textit{u}_1$. $\textbf{R}1(\textbf{q})$ has full rank and is therefore inversible for all admissible states.
The left eigenvector matrix reads 

\footnotesize
\begin{align}
\mathbf{L}_{1}(\mathbf{q}) & = \frac{1}{c^{2}}
\left[\begin{array}{ccccccccc}
\frac{\phi_{1}+u_{1} c}{2} & & & \ldots & \frac{\phi_{K}+u_{1} c}{2} & -\frac{\bar{\gamma} u_{1}+c}{2} & -\frac{\bar{\gamma}}{2} u_{2} & -\frac{\bar{\gamma}}{2} u_{3} & \bar{\gamma} \\
c^{2}-Y_{1} \phi_{1} & -Y_{1} \phi_{2} & & \ldots & -Y_{1} \phi_{K} & Y_{1} \bar{\gamma} u_{1} & Y_{1} \bar{\gamma} u_{2} & Y_{1} \bar{\gamma} u_{3} & -Y_{1} \bar{\gamma} \\
-Y_{2} \phi_{1} & c^{2}-Y_{2} \phi_{2} & \ldots & -Y_{2} \phi_{K} & & & & \\
\vdots & & \ddots & & \vdots & \vdots & \vdots & \vdots & \vdots \\
-Y_{K-1} \phi_{1} & \ldots & c^{2}-Y_{K-1} \phi_{K-1} & -Y_{K-1} \phi_{K} & & & & \\
-Y_{K} \phi_{1} & \ldots & -Y_{K} \phi_{K-1} & c^{2}-Y_{K} \phi_{K} & Y_{K} \bar{\gamma} u_{1} & Y_{K} \bar{\gamma} u_{2} & Y_{K} \bar{\gamma} u_{3} & -Y_{K} \bar{\gamma} \\
-u_{2} c^{2} & & \ldots & -u_{2} c^{2} & 0 & c^{2} & 0 & 0 \\
-u_{3} c^{2} & & \ldots & -u_{3} c^{2} & 0 & 0 & c^{2} & 0 \\
\frac{\phi_{1}-u_{1} c}{2} & & \ldots & \frac{\phi_{K}-u_{1} c}{2} & -\frac{\bar{\gamma} u_{1}-c}{2} & -\frac{\bar{\gamma}}{2} u_{2} & -\frac{\bar{\gamma}}{2} u_{3} & \bar{\gamma}
\end{array}\right]
\end{align}

with \begin{align}
\phi_{i} := \frac{\partial p}{\partial \rho_{i}}= \frac{R}{R-c_{p}}\left(h_{i}-\frac{\mathbf{u}^{2}}{2}\right)-\frac{c_{p}}{R-c_{p}} R_{i} T
= \bar{\gamma}\left(\frac{\mathbf{u}^{2}}{2}-h_{i}\right)+\gamma R_{i} T, \quad \hat{\gamma}=\gamma-1 .
\end{align}

\section{Procedures for characteristic decomposition applied to the MUSCL reconstruction}
\label{apB}
The characteristic decomposition procedures applied here are those proposed in \cite{jiang1996efficient}, only with a small modification.
First, the characteristic variables at cell $j$ are obtained with conservative variables approximated onto the characteristic space by 
\begin{equation}
\boldsymbol{W}_{j} = \boldsymbol{L}_{i} \boldsymbol{Q}_{j}, \quad j = i-1, i, i+1.
\end{equation}
Then the slope of characteristic variables at cell $i$ is given by
\begin{align}
\Delta \boldsymbol{W}_{i} & = \Phi\left(\Delta \boldsymbol{W}_{i-1 / 2}, \Delta \boldsymbol{W}_{i+1 / 2}\right),
\end{align}
where $\Phi$ represents the slope limiter, with $\Delta \boldsymbol{W}_{i+1 / 2}=\boldsymbol{W}_{i+1}-\boldsymbol{W}_{i}$. 

The slope obtained in each characteristic field can then be projected back to the component space by
\begin{eqnarray}
\Delta \boldsymbol{Q}_{i}= \boldsymbol{R}_{i} \Delta W_{i}.
\end{eqnarray}
Finally, the left and right values at the interface can be acquired by
\begin{eqnarray}
\boldsymbol{Q}_{i+1 / 2}^{L}=\boldsymbol{Q}^{l}_{i}+\frac{1}{2} \Delta \boldsymbol{Q}_{i}, \\
\boldsymbol{Q}_{i-1 / 2}^{R}=\boldsymbol{Q}^{l}_{i}-\frac{1}{2} \Delta \boldsymbol{Q}_{i}.
\end{eqnarray}
Here, $L_i$ and $R_i$ are the left and right eigenvector matrices of the Jacobian matrix $A_i$.
This MUSCL scheme is able to maintain good robustness behind the shock discontinuity with relatively low computational cost.



\bibliography{references}



\end{document}